\documentclass[aps,prl,groupedaddress,superscriptaddress,reprint,amsmath,amssymb,showpacs]{revtex4-1}
\usepackage{natbib}
\usepackage{amssymb}
\usepackage{amsmath}
\usepackage{graphicx}
\usepackage[normalem]{ulem}
\usepackage[usenames,dvipsnames]{color}
\usepackage[nointegrals]{wasysym}
\usepackage{siunitx}
\usepackage{dcolumn}
\usepackage{blkarray}
\usepackage{multirow}
\include{commands}

\DeclareRobustCommand{\QD}{\includegraphics{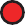}}

\definecolor{triangle}{cmyk}{0.61,0,0.22,0}
\definecolor{square}{gray}{0.5}
\definecolor{lowP}{cmyk}{0.15,0.78,1,0.04}
\definecolor{highP}{cmyk}{0.61,1,0.14,0.03}
\definecolor{5mT}{cmyk}{0,0.61,0.94,0}
\definecolor{25mT}{cmyk}{0.21,1,1,0.18}
\definecolor{45mT}{cmyk}{0.33,0.78,0,0}
\definecolor{gray}{rgb}{.6,.6,.6}
\definecolor{darkyellow}{rgb}{.6,.5,0}
\definecolor{darkgreen}{rgb}{0,.6,0}

\newcommand{\uu}{ \uparrow \uparrow }
\newcommand{\dd}{ \downarrow \downarrow }
\newcommand{\ud}{ \uparrow \downarrow }
\newcommand{\du}{ \downarrow \uparrow }

\newcommand{\abs}[1]{\left\vert #1 \right\vert} 
\newcommand{\avg}[1]{\left< #1 \right>} 
\newcommand{\bra}[1]{\left\langle #1 \right\vert} 
\newcommand{\ket}[1]{\left\vert #1 \right\rangle} 

\newcommand{\braket}[2]{\bra{#1 \vphantom{#2}} \left. #2 \vphantom{#1} \right\rangle}

\newcommand{\ketbra}[2]{\ket{#1 \vphantom{#2}} \bra{#2 \vphantom{#1}}}

\newcommand{\matrixel}[3]{\left< #1 \vphantom{#2#3} \right| #2 \left| #3 \vphantom{#1#2} \right>}

\newcommand{\Ez}{\ensuremath{E_\text{z}}}
\newcommand{\Bext}{\ensuremath{\mathbf{B}_\text{ext}}}

\newcommand{\Bm}{\ensuremath{\mathbf{B}_\text{nm}}}

\newcommand{\Bmd}{\ensuremath{\Delta \mathbf{B}_\text{nm}}}
\newcommand{\Bml}{\ensuremath{\mathbf{B}_\text{nm}^\text{L}}}
\newcommand{\Bmr}{\ensuremath{\mathbf{B}_\text{nm}^\text{R}}}
\newcommand{\Bn}{\ensuremath{\mathbf{B}_\text{nuc}}}
\newcommand{\Bnz}{\ensuremath{B_\text{nuc}^z}}
\newcommand{\Bnmax}{\ensuremath{B^\text{max}_\text{nuc}}}

\newcommand{\Bnd}{\ensuremath{\Delta \mathbf{B}_\text{nuc}}}
\newcommand{\Bd}{\ensuremath{\Delta \mathbf{B}}}
\newcommand{\Sd}{\ensuremath{\Delta \mathrm{S}}}

\newcommand{\Bs}{\ensuremath{\overline{\mathbf{B}}}}
\newcommand{\Ss}{\ensuremath{\overline{\mathrm{S}}}}

\newcommand{\Bl}{\ensuremath{\mathbf{B}^\text{L}}}
\newcommand{\Br}{\ensuremath{\mathbf{B}^\text{R}}}
\newcommand{\Sl}{\ensuremath{\mathbf{S}^\text{L}}}
\newcommand{\Sr}{\ensuremath{\mathbf{S}^\text{R}}}

\newcommand{\Tp}{\ensuremath{T_+}}
\newcommand{\Tm}{\ensuremath{T_-}}
\newcommand{\Sp}{\ensuremath{S_\mathrm s}}
\newcommand{\Sm}{\ensuremath{S_\mathrm a}}
\newcommand{\Tzero}{\ensuremath{T_0}}

\newcommand{\Sa}{\ensuremath{S_\text{a}}}
\newcommand{\Slr}{\ensuremath{S_{11}}}
\newcommand{\Srr}{\ensuremath{S_{02}}}
\newcommand{\tc}{\ensuremath{t_\text{c}}}
\newcommand{\Del}{\ensuremath{\Delta}}
\newcommand{\mB}{\ensuremath{\mu_\text{B}}}
\newcommand{\Imax}{\ensuremath{I_\text{max}}}

\newcommand{\Grel}{\ensuremath{\Gamma_\text{rel}}}
\newcommand{\Gpol}{\ensuremath{\Gamma_\text{pol}}}
\newcommand{\Gpolmax}{\ensuremath{\Gamma_\text{pol}^\text{max}}}
\newcommand{\Pmax}{\ensuremath{P_\text{max}}}
\newcommand{\dP}{\ensuremath{\dot{P}}}
\newcommand{\dPmax}{\ensuremath{\dot{P}_\text{max}}}

\newcommand{\proj}[1]{\ketbra{#1}{#1}}

\newcommand{\nocontentsline}[3]{}
\newcommand{\tocless}[2]{\bgroup\let\addcontentsline=\nocontentsline#1{#2}\egroup}

\begin{document}

\title{Large nuclear spin polarization in gate-defined quantum dots using a single-domain nanomagnet}

\author{Gunnar Petersen}
\thanks{These authors contributed equally to this work.}
\affiliation{Center for Nanoscience and Fakult\"{a}t f\"{u}r Physik, Ludwig-Maximilians-Universit\"{a}t M\"{u}nchen, \\
Geschwister-Scholl-Platz 1, 80539 M\"{u}nchen, Germany}

\author{Eric~A. Hoffmann}
\thanks{These authors contributed equally to this work.}
\affiliation{Center for Nanoscience and Fakult\"{a}t f\"{u}r Physik, Ludwig-Maximilians-Universit\"{a}t M\"{u}nchen, \\
Geschwister-Scholl-Platz 1, 80539 M\"{u}nchen, Germany}

\author{Dieter Schuh}
\affiliation{Institut f\"ur Experimentelle Physik, Universit\"at Regensburg, D-93040 Regensburg, Germany}

\author{Werner Wegscheider}
\affiliation{Solid State Physics Laboratory, ETH Zurich, 8093 Zurich, Switzerland}
\affiliation{Institut f\"ur Experimentelle Physik, Universit\"at Regensburg, D-93040 Regensburg, Germany}

\author{Geza Giedke}
\affiliation{Max-Planck-Institut f\"{u}r Quantenoptik, 85748 Garching, Germany}

\author{Stefan Ludwig}
\affiliation{Center for Nanoscience and Fakult\"{a}t f\"{u}r Physik, Ludwig-Maximilians-Universit\"{a}t M\"{u}nchen, \\
Geschwister-Scholl-Platz 1, 80539 M\"{u}nchen, Germany}
\email{ludwig@lmu.de}
\date{\today}




\begin{abstract}
The electron-nuclei (hyperfine) interaction is central to spin qubits in solid state systems. It can be a severe decoherence source but also
allows dynamic access to the nuclear spin states. We study a double
quantum dot exposed to an on-chip single-domain nanomagnet and show that
its inhomogeneous magnetic field crucially modifies
the complex nuclear spin dynamics such that the Overhauser field tends
to compensate external magnetic fields. This turns out to be beneficial for polarizing the nuclear spin ensemble. We reach a nuclear spin polarization of $\simeq \SI{50}{\percent}$, unrivaled in lateral dots, and explain our manipulation technique using a comprehensive rate equation model.
\end{abstract}

\maketitle

The hyperfine interaction (HFI) between  few electrons and a bath of nuclear
spins induces a complex quantum many-body dynamics which has been studied in a variety of systems including  phosphorus donors in silicon \cite{Kane98}\@,
nitrogen vacancy centers in diamond \cite{Jelezko06}\@, quantum Hall
systems \cite{Kou10}\@, and semiconductor-based quantum dots, both optically  \cite{Urbaszek12} and in transport \cite{Schliemann03,Fischer09}\@. In GaAs double quantum dots (DQDs), each electron interacts with \mbox{$\sim$$10^6$} nuclear spins, which fluctuate thermally even at cryogenic temperatures. Their HFI causes  electron spin decoherence \cite{Yao06,Fischer09,Cywinski09,Bluhm11}\@, but it also offers a means to control the nuclear spins dynamically \cite{Laird07,Petta08,Foletti09,Reilly10}\@. As has been proposed \cite{Burkard99,Klauser06,Rudner07,Danon08} and demonstrated \cite{Latta09,Vink09,Danon09}\@, nuclear spin manipulation facilitates nuclear state preparation, which can enhance spin coherence times \cite{Bluhm10,Sun12}\@. Nuclear spin manipulation has also motivated theoretical proposals \cite{Taylor03,Kurucz09} and experimental realizations \cite{Maurer12,Steger12} of nuclear spins as quantum memory.

Here, we present a novel nuclear spin manipulation technique. We couple a DQD with a single-domain nanomagnet whose inhomogeneous magnetic field does not depend on external fields. This allows measurements in a new regime, which will increase our fundamental understanding and control of the coupled electron-nuclei system common to a variety of platforms. As an example, we demonstrate unusually strong dynamic nuclear spin polarization (DNSP).
In laterally defined DQDs, adiabatic pumping experiments have produced polarizations of \SIrange{1}{5}{\percent} \cite{Petta08,Foletti09,Reilly10}\@, and using electron dipole spin resonance, \SI{16}{\percent} has been realized \cite{Laird07}\@. Using a simpler experimental technique, we report $\simeq \SI{50}{\percent}$ polarization, achieved by exploiting the benefits of our nanomagnet.

\begin{figure}[t!]
\begin{center}
\includegraphics{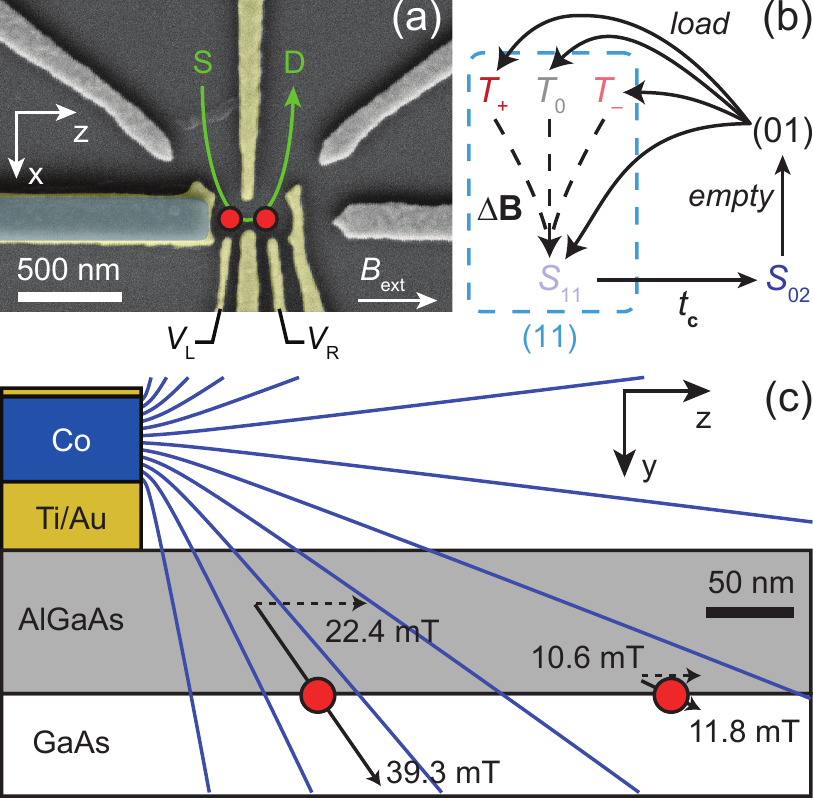}
\caption{
(a) Scanning electron micrograph showing the Ti/Au gates (yellow) used to define the DQD. Electrons tunnel sequentially from source (S) to drain (D) through the two QDs located near $\QD$.
The (blue) single-domain Co nanomagnet (200 (width) $\times$ 50 (height) $\times$ \SI{2000}{\nano\meter\cubed}, $2/5$ visible) generates an inhomogeneous magnetic field, \Bm.
(b)
The tunneling sequence $(0\,1) \to (1\,1) \to (0\,2) \to (0\,1)$. Triplet-singlet transitions $\{ \Tm, \Tzero, \Tp \} \to \Slr$ require lifting of the PSB.
(c) The relevant layers of the heterostructure, gates, and nanomagnet. The magnetic field lines (blue) are calculated according to ref.~\citenum{Engel05}\@.
The black arrows indicate the local magnetic fields $\Bm^\text{L} = \SIlist{0;32.3;22.4}{\milli\tesla}$ and $\Bm^\text{R} = \SIlist{0;5.2;10.6}{\milli\tesla}$. The \SI{3}{\milli\tesla} radius of $\QD$ represents the typical Overhauser field fluctuations.
\label{fig:setup}
}
\end{center}
\end{figure}

\begin{figure}[t!]
\begin{center}
\includegraphics{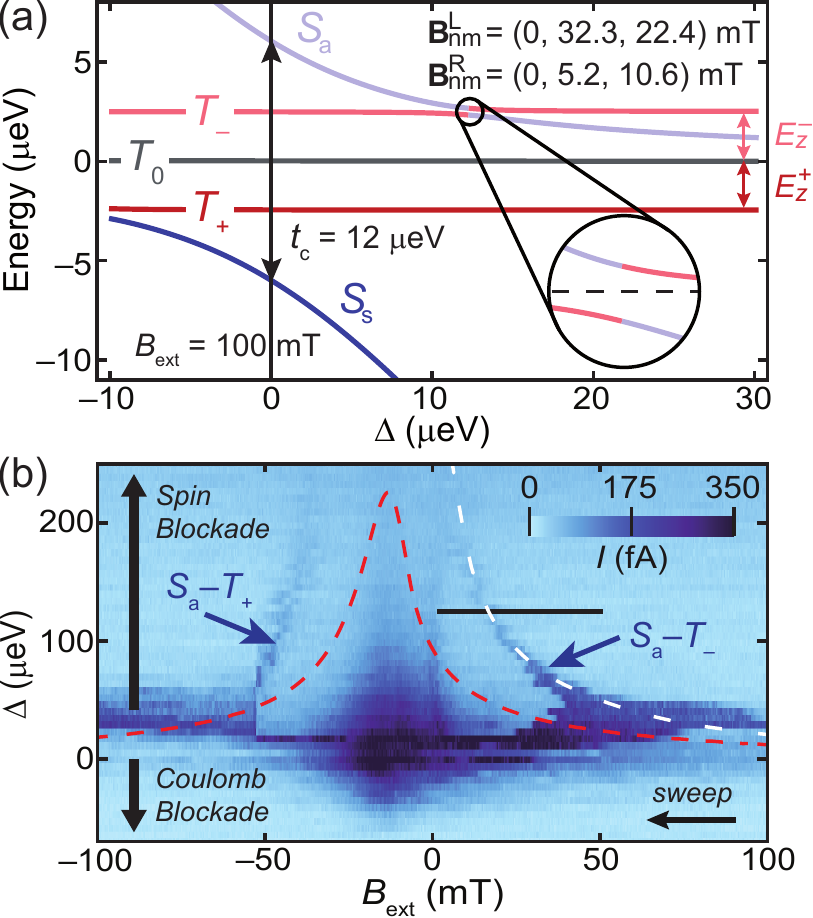}
\caption{
(a) The relevant eigenenergies as a function of $\Delta$, for $\Bn = 0$. The inhomogeneous \Bm\ causes singlet-triplet mixing (see enlarged $\Sa$--$T_-$ avoided crossing) and lifts the PSB.
(b) Leakage current, $I$, through the DQD versus \Del\ (stepped from top to bottom) and $B_\text{ext}$ (swept at $-50$\,mT$/$min). The red dashed line is a numerical prediction of the $\Sa$--$T_\pm$ resonances without DNSP using $\tc = \SI{12}{\micro\electronvolt}$. The white dashed line includes DNSP (via the $\Sm$--$\Tm$ resonance) using eq~\ref{eq:DNSP} with $\Grel = \SI{0.043}{\per\second}$, $\gamma = \SI{10}{\milli\tesla}$, and $\alpha = \SI{35}{\per\nano\ampere\per\second}$.
\label{fig:Energy and B vs Delta}
}
\end{center}
\end{figure}

We measure the current, $I$, which results from a dc voltage, here $V = 1$\,mV, applied across the DQD exposed to the inhomogeneous magnetic field of a nanomagnet (see Figures~\ref{fig:setup}a,c). As detailed in Figure~\ref{fig:setup}b, electrons tunnel sequentially through the DQD via the occupation cycle $(0\,1) \rightarrow (1\,1) \rightarrow (0\,2) \rightarrow (0\,1)$, where $(m\,n)$ indicates the number, $m$ ($n$), of electrons in the left (right) dot. The transition $(0\,1) \rightarrow (1\,1)$ loads one of four (1\,1) states, which consist of the singlet state, $S_{11}$, and the three triplet states, $T_{11} = \left\lbrace T_-,~T_0,~T_+ \right\rbrace$. Only $T_\pm$ have a nonzero spin projection along the quantization axis, which we choose parallel to the external magnetic field, $B_\text{ext}$ (along the $z$-axis in Figure~\ref{fig:setup}a). The only energetically accessible (0\,2) state is the singlet state, $S_{02}$. The corresponding eigenenergies are plotted in Figure\ \ref{fig:Energy and B vs Delta}a as a function of the energy detuning, $\Delta$, between the diabatic singlet states \Slr\ and $\Srr$. The singlet eigenstates, \Sp\ and \Sm, are the symmetric and antisymmetric combinations of \Slr\ and \Srr\ mixed by the interdot tunnel coupling, \tc. $T_0$ is at zero energy (neglecting exchange coupling), while $T_\pm$ are split by their Zeeman energies, $E^\pm_\text{z} = \pm g \mB \left( \abs{\Bl + \Br} \right)/2$, where \Bl\ (\Br) is the local magnetic field in the left (right) dot, $g$ the Land\'{e} g-factor, and $\mB$ the Bohr magneton. For $\Del > 0$, the triplets are well separated from $\Sp$, but not from $\Sa$.

In a homogeneous magnetic field $\left( \Bl = \Br \right)$, transitions
between triplets and singlets are forbidden by Pauli spin blockade \cite{Ono02}
(PSB) (dashed arrows in Figure~\ref{fig:setup}b). Eventually the occupation
cycle stalls in one of the triplets resulting in $I = 0$ (neglecting
cotunneling). DNSP requires $I\ne0$, which can be induced by a local field difference, $\Bl - \Br$, that lifts the PSB by coupling triplets to singlets. One way to produce inhomogeneous fields is to include on-chip micromagnets, which have been used for all-electric control of a single electron spin \cite{Pioro08,Obata10,Brunner11}\@. However, at external fields below a few hundred mT, micromagnets form magnetic domains, which greatly reduce their fields. Here, we use a nanomagnet (see Figure\ \ref{fig:setup}a), which forms a single magnetic domain (due to its shape anisotropy) and a sizable $\Bd = \left( \Bl - \Br \right) / 2$ even if $B_\text{ext} = 0$ \cite{Heedt12}\@. This previously unexplored regime proves to be highly beneficial for controlling DNSP.

Even in the absence of on-chip magnets, the HFI between thermally fluctuating nuclear spins and electrons creates an \emph{effective} (Overhauser) field, $\Bn$, which statistically varies between the two dots resulting in a small leakage current near $B_\text{ext} \simeq 0$ and $\Del \simeq 0$ \cite{Suppl,Koppens05}. Compared to $\Bn$, the nanomagnet's inhomogeneous field, $\Bm$, is more stable in time, and $\abs{\Bmd} \gg \abs{\Bnd}$ (see Figure~\ref{fig:setup}c). $B_\text{ext}$ is aligned along the easy axis (z-axis) of the nanomagnet, which has a coercive field of \SI{52}{\milli\tesla}. Because the nanomagnet forms a single domain, $B_\text{ext}$ does not affect the magnitude of \Bm. The associated $\Bmd$ results in a leakage current over a broad range of \Del\ and $B_\text{ext}$ including distinct current maxima along the \Sm--$T_\pm$ resonances (Figure~\ref{fig:Energy and B vs Delta}b). These current features are seen at small $B_\text{ext}$ and, hence, are not accessible
with multidomain magnets (see above). Our observed current features are very different from measurements performed without an on-chip magnet \cite{Suppl}.

However, consideration of $\Bm$ alone does not explain all features in Figure~\ref{fig:Energy and B vs Delta}b. We must include the HFI and its effect on DNSP, as has proven necessary in other experiments \cite{Ono04,Vink09,Kobayashi11,Frolov12}\@. The dashed red line in Figure~\ref{fig:Energy and B vs Delta}b is a prediction of the position of the $\Sa$--$T_\pm$ resonances as a function of $B_\text{ext}$\ and \Del. It takes into account $\Bm$, but neglects $\Bn$ \cite{Suppl}. Compared to this prediction, however, the measured resonances (blue arrows in Figure~\ref{fig:Energy and B vs Delta}b) occur at larger $\abs{B_\text{ext}}$. As we will show, this shift can be explained by including DNSP, which produces a $\Bnz$ that compensates $B_\text{ext}$, e.\,g., $\Bnz < 0$ when $B_\text{ext}>0$.

\begin{figure}[t!]
\begin{center}
\includegraphics{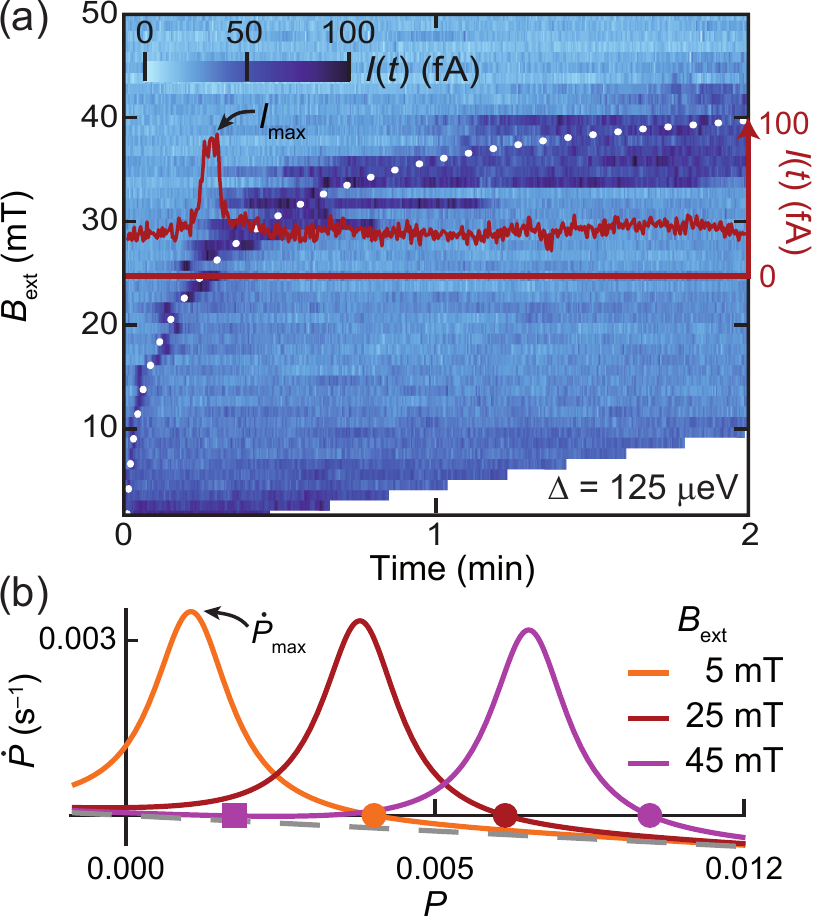}
\caption{
(a) Current versus time,  $I(t)$, as a function of $B_\text{ext}$ along the horizontal line in Figure~\ref{fig:Energy and B vs Delta}b {(we prepared $P(0) = 0$).}
\emph{Density Plot: $I(t,B_\text{ext})$.} The white dotted line predicts the moment of maximal current, $t\left( \Imax \right)$, using eq~\ref{eq:DNSP} and the same parameters as in Figure~\ref{fig:Energy and B vs Delta}b.
\emph{Red Trace:} $I(t)$ measured at $B_\text{ext} = \SI{25}{\milli\tesla}$.
(b) $\dP(P)$ for three different $B_\text{ext}$ using eq \ref{eq:DNSP} and the same parameters as in Figure~\ref{fig:Energy and B vs Delta}b. Far from resonance an exponential decay $\dP=-\Grel P$ remains (gray dashed line). {\color{5mT}$\CIRCLE$},{\color{25mT}$\CIRCLE$},{\color{45mT}$\CIRCLE$} mark the $B_\text{ext}$--dependent ``adjustable'' fixed point and {\color{45mT}$\blacksquare$} a ``trivial'' fixed point, which appears near $P = 0$ when $B_\text{ext} \gtrsim \SI{45}{\milli\tesla}$.
\label{fig:timeevolution}
}
\end{center}
\end{figure}

The connection to DNSP becomes evident with the data shown in Figure~\ref{fig:timeevolution}a, which probes the \Sm--\Tm\ resonance as a function of time for a fixed \Del. We prepared the nuclear spin polarization, $P$, to $P \simeq 0$ by waiting three minutes at $I = 0$ before turning on the voltage across the DQD. The current maximum, $\Imax$, at the \Sm--\Tm\ resonance occurs later in time at larger $B_\text{ext}$. Again, this can be explained if $B^z_\text{nuc} < 0$ and compensates $B_\text{ext}$. Because the GaAs g-factor is negative, $P = -\Bnz / B^\text{max}_\text{nuc}$, where $B^\text{max}_\text{nuc} \simeq 6.1$\,T is the Overhauser field magnitude produced when all nuclear spins are aligned \cite{Suppl}. $B^z_\text{nuc} < 0$ implies that $P > 0$, which can only be explained if DNSP from \Tp\ outweighs that from $\Tm$, despite the system being near the $\Sa$--$T_-$ resonance. This peculiar situation results from spin-selective lifting of the PSB (of $T_-$) and bolsters DNSP, as discussed
below.

Our explanation starts with a rate equation model \cite{Vink09,Danon09,Rudner11} including only polarization generated by $T_+$ near the $\Sa$--$T_-$ resonance (A comprehensive calculation in \cite{Suppl} Sec. III includes all (1\,1) states). As a simplification, we use the average polarization $P = \left( P^\text{L} + P^\text{R} \right) / 2$. ($P^\text{L} \neq P^\text{R}$ would mainly affect the decay of $T_0$, not $T_\pm$.) The overall rate equation is
\begin{equation}
\dP(t)=\Gpol(t)\left[ 1 - P(t) \right] - \Grel P(t),
\label{eq:DNSP}
\end{equation}
where the polarization decay rate, $\Grel$, is constant, while the build-up rate, $\Gpol(t)$, is proportional to current, $\Gpol(t) = \alpha I(t) > 0$, as observed experimentally. For convenience, we describe the current maximum at the \Sm--\Tm\ resonance as a Lorentzian
\begin{equation}
I(t) = \Imax \frac{\left( \gamma/2 \right)^2}{\left(E^-_z - E_\text{a} \right)^2 + \left( \gamma/2 \right)^2},
\label{eq:lorentz}
\end{equation}
where \Imax\ is the (measured) resonant current and $\gamma$ is the effective width of the resonance. (Nonresonant states contribute weakly to $I(t)$.) Here
\begin{equation}
E_\text{a} = \left( -\Del + \sqrt{\tc^2+\Del^2} \right)/2
\label{eq:Ea}
\end{equation}
is the energy of $\Sa$. We approximate $E_\text{z}$ by only including the average $z$-component of $\Bm$, $\overline{B}^z_\text{nm}$, so that
\begin{equation}
E^-_\text{z} \approx \abs{g} \mB \left( B_\text{ext}(t) - \Bnmax P(t) + \overline{B}^z_\text{nm} \right).
\label{eq:Ez}
\end{equation}

Example $\dP(P)$ curves are plotted in Figure~\ref{fig:timeevolution}b and are used to model the data in Figure\ \ref{fig:timeevolution}a. $P\left( t = 0 \right) = 0$ in this measurement, and the model predicts $\dP(P = 0) > 0$ (evident in Figure~\ref{fig:timeevolution}b). Therefore, $P$ increases in time until it reaches a stable fixed point at $\dP = 0$ (and $\text d\dP/\text d P < 0$). For $B_\text{ext} < \SI{43}{\milli\tesla}$, $P$ passes through the $\Sm$--$\Tm$ resonance, which coincides with $\dPmax$ in Figure~\ref{fig:timeevolution}b. As $B_\text{ext}$ is increased, the \Sm--\Tm\ resonance moves to larger $P$, and with it move $\dPmax$ and the stable ``adjustable'' fixed point (A-FP, circles in Figure\ \ref{fig:timeevolution}b). Accordingly, the measured (resonant) $\Imax$ in Figures\ \ref{fig:timeevolution}a appears later in time with increasing $B_\text{ext}$. When $B_\text{ext} \approx \SI{45}{\milli\tesla}$, a second stable ``trivial'' fixed point (T-FP, square in Figure\ \ref{fig:timeevolution}b) appears near $P = 0$ and remains there for
$B_\text{ext} > \SI{45}{\milli\tesla}$. Hence, we expect $P$ to remain near zero (far from resonance) at the T-FP. Indeed, no resonant current maximum is observed for $B_\text{ext} \gtrsim \SI{43}{\milli\tesla}$ in Figure\ \ref{fig:timeevolution}a.

Eq \ref{eq:DNSP} provides quantitative predictions of the time evolution of
the \Sm--\Tm\ resonance associated with the measured $\Imax$. Namely, it
yields the white fits in Figures\ \ref{fig:Energy and B vs Delta}b and
\ref{fig:timeevolution}a. These two separate fits share altogether four fit
parameters. The agreement between our model and data indicates that the model captures the DNSP in both experiments. In addition, $\Grel$ agrees with reported values \cite{Reilly10}\@.

\begin{figure}[t!]
\begin{center}
\includegraphics{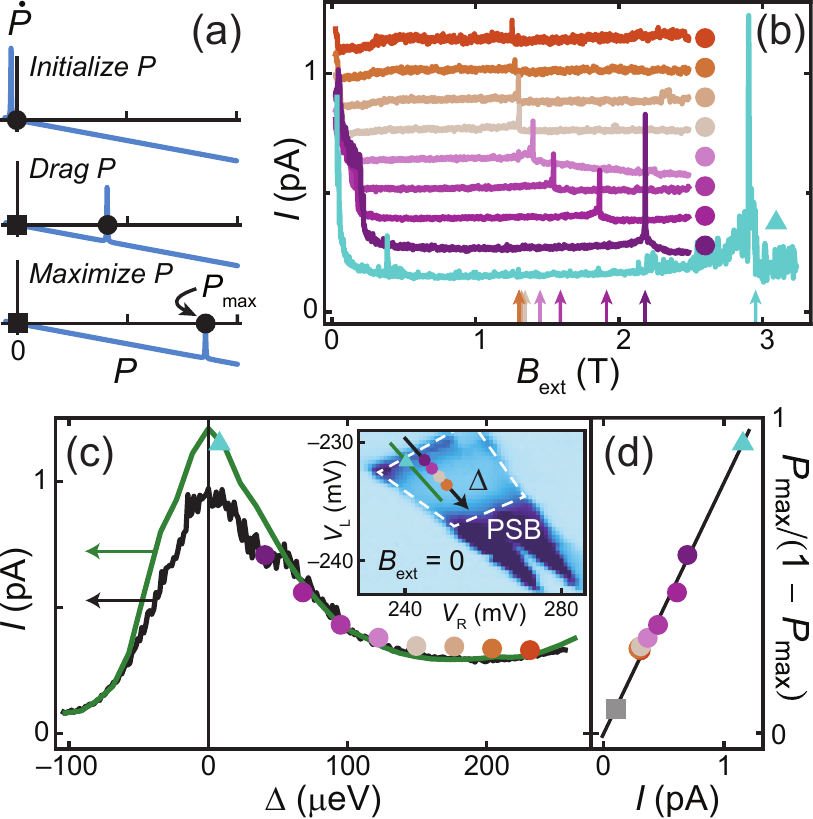}
\caption{
(a) $\dot{P}(P)$ for three $B_\text{ext}$ using eq \ref{eq:DNSP} with $\Imax = \SI{1.2}{\pico\ampere}$ and the fit parameters from Figure~\ref{fig:Energy and B vs Delta}b. $\blacksquare$ marks the T-FP and $\CIRCLE$ the A-FP.
(b) Leakage current, $I$, for various $\Del$ in the PSB regime [see (c)] measured while continuously increasing $B_\text{ext}$ (at 30\,mT$/$min) after initialization at the A-FP at $B_\text{ext} = \SI{50}{\milli\tesla}$. (All but the lowest trace are vertically offset in increments of 125\,fA.) Sharp current peaks (marked by arrows) correspond to the maximum polarization, $\Pmax(\Del)$, in (a).
(c) $P_\text{max}/\left( 1 - P_\text{max} \right)$ from (b) (filled circles) and $I$ (black/green traces), measured along the (black/green) lines in the inset, versus $\Delta$. \textit{Inset}: $I$ measured as a function of gate voltages $V_\text{L,R}$ (see Figure\ \ref{fig:setup}a). The region of suppressed $I$ within the double triangle of finite current marks PSB.
(d) $P_\text{max}/\left( 1 - P_\text{max} \right)$ versus $I$, extracted from the traces in (c). The gray square is predicted from the data in Figure~\ref{fig:Energy and B vs Delta}b. The black line expresses eq \ref{eq:P max} with  $\alpha / \Grel = \SI{0.81}{\per\pico\ampere}$ using the fit parameters from Figure~\ref{fig:Energy and B vs Delta}b.
\label{fig:highpol}
}
\end{center}
\end{figure}

Our model reveals a straightforward procedure to maximize $P$. We start at small $B_\text{ext}$ where the T-FP is absent and $P$ is initialized at the A-FP (see top panel of Figure \ref{fig:highpol}a). This initialization requires small $B_\text{ext}$ and sufficient singlet--triplet mixing and is only possible with a single-domain nanomagnet. To reach a large $P$, we increase $B_\text{ext}$ (with a sufficiently slow sweep rate) dragging the A-FP, and $P$ along with it (see middle panel of Figure \ref{fig:highpol}a). This dragging procedure works up to a maximum polarization, $\Pmax$,  occurring when the decay of $P$ overwhelms its build-up and $\dPmax = 0$ (see bottom panel of Figure \ref{fig:highpol}a). $\Pmax$ is defined by solving eq~\ref{eq:DNSP} for $\dP(\Pmax) = 0$:
\begin{equation}
\Pmax / \left( 1 - \Pmax \right) =  \alpha \Imax / \Grel \equiv \Gpolmax / \Grel,
\label{eq:P max}
\end{equation}
where $\Gpolmax, \Pmax > 0$ and both depend on $\Delta$ via $\Imax$. At $B^\text{max}_\text{ext}$ (the field corresponding to $\Pmax$), the A-FP coincides with the $\Sa$--$T_-$ resonance, and we expect a current maximum.

Indeed, the leakage current measured during such DNSP sweeps, displayed for various $\Delta$ in Figure \ref{fig:highpol}b, features a sharp current maximum at a particular $B_\text{ext}$ (see arrows),
which we identify as $B^\text{max}_\text{ext}(\Del)$. For $B_\text{ext} > B^\text{max}_\text{ext}$, the A-FP is lost, and the nuclear spin polarization decays ($\dP < 0$) with rate \Grel.

$P_\text{max}(\Del)$ compensates $B^\text{max}_\text{ext}(\Del)$ so that we can equate $P_\text{max}(\Del) =  B^\text{max}_\text{ext}(\Del)/B^\text{max}_\text{nuc}$ . $\Pmax/(1 - \Pmax)$ and $I$ measured versus $\Delta$ are shown side-by-side in Figure\
\ref{fig:highpol}c demonstrating a striking correlation between $\Pmax/(1 - \Pmax)$ and $I$. In fact, in accordance with eq \ref{eq:P max}, Figure~\ref{fig:highpol}d shows that $\Pmax / \left( 1 - \Pmax \right) \propto I$ confirming our assumption that $\Gpol \propto I$. The straight black line
in Figure~\ref{fig:highpol}d expresses eq~\ref{eq:P
  max} using the fit parameters from Figure \ref{fig:Energy and B vs Delta}b. The ability to explain three very different data
sets (Figures \ref{fig:Energy and B vs Delta}b, \ref{fig:timeevolution}a,
\ref{fig:highpol}d) with one set of fit parameters corroborates the
interpretation of the current peaks in Figure~\ref{fig:highpol}b and the
validity of our rate equation model \cite{Suppl}\@.

Our highest $B^\text{max}_\text{ext} \simeq \SI{2.9}{\tesla}$ (green data in Figure\ \ref{fig:highpol}b) corresponds to $P \simeq \SI{50}{\percent}$ (and generates an Overhauser field gradient of $\sim \SI[per-mode=symbol]{1}{\tesla\per{100\nano\meter}}$ across the DQD boundary). This exceeds by far previously reported polarizations in laterally defined DQDs \cite{Laird07,Petta08,Foletti09,Reilly10}\@(a complementary measurement of P in \cite{Suppl} Sec. VI).

To detail how $\Bm$ and the HFI combine to lift the PSB and induce DNSP, we compare our system with two simpler scenarios. If the HFI were the only mechanism to lift PSB, no DNSP would be expected since all triplets are loaded equally often resulting in as many up as down nuclear spin flips. In experiments without a nanomagnet \cite{Baugh07,Kobayashi11}\@, cotunneling weakly lifts the PSB (in competition with the HFI) and does so equally for each triplet, nearly irrespective of its energy. In contrast, the hyperfine-induced decay rate is strongly energy dependent. Therefore, near the $\Sa$--$T_-$ resonance, $T_-$ generates more nuclear spin flips than $T_+$, and $P < 0$ is observed without nanomagnet.

In our case, $\Bm$ mixes \Tm\ and \Sm\ strongly near their resonance, resulting in two (1\,1) states that are no longer in PSB. Hyperfine-induced decay is heavily suppressed in these mixed states. In this situation, the HFI still contributes to the decay of $T_+$ (and $T_0$) thereby producing DNSP and $P > 0$.
In an alternative approach, DNSP has been studied for large $\tc$ ($\sim \SI{100}{\micro\electronvolt}$) by sweeping \Del \cite{Baugh07,Kobayashi11,Frolov12}\@.
However, when $\tc \sim \SIrange{1}{10}{\micro\electronvolt}$, which is favorable for spin qubits, $P$ is limited by the energy of $\Sa$ in the PSB regime, so that $\abs{P} \lesssim \tc/\left( 2 \abs{g} \mB \Bnmax \right) \sim \SI{10}{\percent}$. Moreover, $\Bnz < 0$ in our system provides a distinct advantage because $\Bnz$ compensates $B_\text{ext}$ such that the total effective magnetic field is constant during the polarization build-up; therefore, $P$ is only limited by $\alpha \Imax/\Grel$ when dragging $P$ with $B_\text{ext}$.

We have demonstrated a nuclear spin polarization of  $\simeq \SI{50}{\percent}$ in a DQD based on the enhanced ability to manipulate the nuclear spin ensemble using an on-chip nanomagnet. Larger polarizations can be expected upon further optimization of the electronic spectrum, sample geometry and materials. Our results demonstrate the flexibility offered by an on-chip nanomagnet, which could be used for all-electric ESR \cite{Pioro08} while simultaneously polarizing the nuclear spin ensemble at small $\tc$ values ideal for spin qubit operation. Such a system could be used to improve nuclear state preparation techniques \cite{Giedke06,Reilly08,Foletti09,Issler10} or for measuring complex nuclear phenomena such as spin squeezing \cite{Rudner11a}\@, quantum memory \cite{Taylor03,Kurucz09}\@, dark states \cite{Gullans10}\@, quantum phase transitions \cite{RudnerPRB10}\@, and superradiance \cite{Schuetz12}\@.

\tocless{\begin{acknowledgments}
We thank S.\ Cammerer as well as J.\,P.\ Kotthaus for helpful discussions regarding the nanomagnet design and S.\ Manus for technical support. Financial support from the German Science Foundation DFG via SFB 631 and the German Excellence Initiative via the ``Nanosystems Initiative Munich (NIM)'' is gratefully acknowledged. E.\,A.\,H.\ thanks the Alexander von Humboldt Foundation and S.\,L.\ the Heisenberg program of the DFG.
\end{acknowledgments}}

\renewcommand{\bibsection}{}

\renewcommand{\thefigure}{S\arabic{figure}}
\renewcommand{\theequation}{S\arabic{equation}}
\renewcommand{\citenumfont}[1]{\text{R#1}}
\setcounter{figure}{0}
\setcounter{equation}{0}
\onecolumngrid
\clearpage
\newpage
\pagestyle{plain}

\tocless\section{Supplementary Information for ``Large nuclear spin polarization in gate-defined quantum dots using a single-domain nanomagnet''}


\tableofcontents

\section{I.	Overview}
\label{sec:overview}
The following supplementary material provides additional information related
to various aspects of the main article. We start in Section II\ref{sec:setup} with details about the sample design and the experimental setup of the measurements. In Section III\ref{sec:hamiltonian} we introduce a model Hamilton operator which describes the hyperfine interaction (HFI) in our double quantum dot (DQD) setup including the inhomogeneous magnetic field of the nanomagnet.
Based on a rate equation model, in Section IV\ref{sec:perturbation}, we perturbatively solve the dynamic nuclear spin polarization (DNSP) problem by explicitly taking into account the contributions of all four (1\,1) states.
We show that the perturbative solution justifies the simplified model used in the main article. Section V\ref{sec:current} provides detailed explanations of the current features in Fig.\ 2b of the main
article. In Section VI\ref{sec:buildup} we present results of a complementary experiment that gives additional evidence of the validity of our model and of our interpretation of the data
in Fig.\ 4 of the main article in terms of a large nuclear spin polarization. Section VII\ref{sec:fitting} describes the fitting procedure for the data in Figs.\ 2b,\ 3a of the main article.
\section{II.	Sample Design and Experimental Setup}
\label{sec:setup}
\begin{figure}[h]
  \includegraphics[width = 0.75\textwidth]{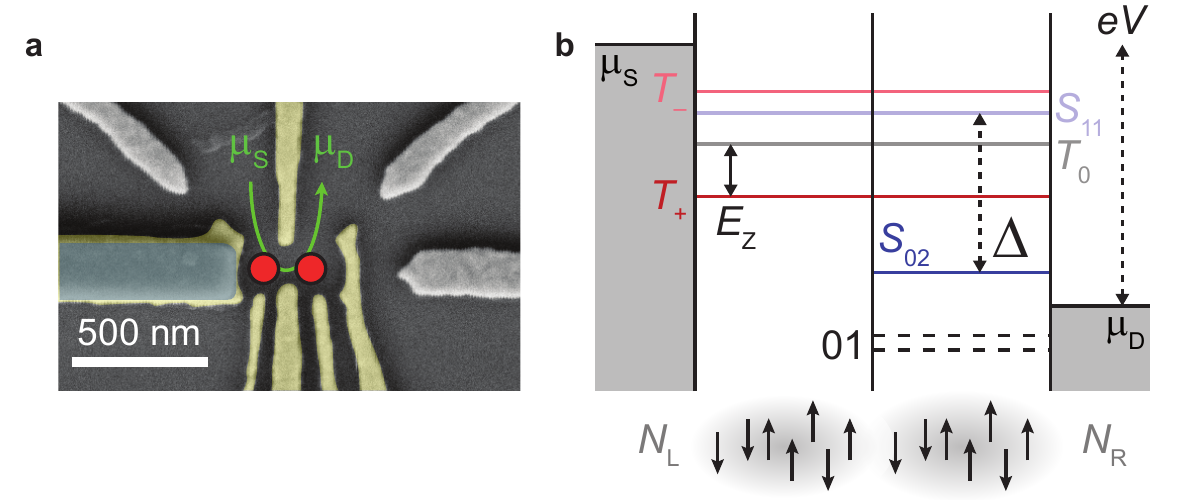}
\caption{\textbf{Experimental setup a)} SEM image of the DQD device used in the experiments.
\textbf{b)} A DQD in Pauli-spin blockade (typical experimental situation), while a voltage $V$ is applied between the degenerate 2D leads with chemical potentials $\mu_\text S$ and $\mu_\text D$. Vertical lines are tunnel barriers. The right dot always contains at least one electron. The dashed horizontal lines are the spin-split chemical potentials of the spin up/down (0\,1) states, where the chemical potential of a quantum dot is defined as the energy needed to add one more electron. The solid horizontal lines are the chemical potentials of the five relevant two-electron basis states, where $E_\text Z$ is the Zeeman energy and \Del\ is the energy detuning between the singlet states, $S_{11}$ and $S_{02}$. 
\label{fig:PSB}
}
\end{figure}

The samples have been fabricated from a GaAs\,/\,AlGaAs heterostructure containing a two-dimensional electron system (2DES) \SI{85}{\nano\meter} below the surface. At cryogenic temperatures, the 2DES has a carrier density of $\SI{1.19e11}{\per\centi\meter\squared}$ and a mobility of $\SI{0.36e6}{\centi\meter\squared\per\volt\per\second}$. Metallic gate electrodes (30\,nm gold on top of 5\,nm titanium) have been fabricated on the sample surface by electron-beam lithography and standard evaporation/lift-off techniques (Fig. ~\ref{fig:PSB}a). The Co nanomagnet with a thickness of \SI{50}{\nano\meter} was evaporated directly on top of the leftmost gate and capped with \SI{5}{\nano\meter} of Au to prevent oxidization. Negative voltages applied to these electrodes are used to deplete locally the 2DES and thereby define the DQD. The absolute electron occupation, ($m~n$), was determined by quantum-point-contact charge detection\cite{R1}. All measurements have been performed in a dilution refrigerator at an electron temperature of $\sim \SI{100}{\milli\kelvin}$. Fig.~\ref{fig:PSB}b sketches the experimental situation in this Letter. A source-drain voltage of $V = (\mu_\text{S} - \mu_\text{D})/e$ is applied across the DQD between degenerate leads. The DQD is in the Pauli-spin blockade, where a triplet state can only contribute to current if it is coupled to a singlet state, e.g., by field inhomogeneity or interaction with the ensemble of $N_\text{L(R)} \sim10^6$ nuclei.

\section{III.	The Hamiltonian}
\label{sec:hamiltonian}
The total Hamiltonian of the system includes electrostatic, magnetic, and
hyperfine contributions and is given (in the relevant subspace depicted in Fig.~\ref{fig:PSB}b) by
\begin{equation}
  H = H_\text{el} + H_\text{B} + H_\text{hf}.
  \label{eq:H}
\end{equation}
Using the diabatic singlet-triplet basis $\left\lbrace
  T_+,T_0,T_-,S_{11},S_{02} \right\rbrace$, the electrostatic contribution is
\begin{equation}
  H_{\text{el}} = t_\text{c}/2 \left( \ketbra{S_{11}}{S_{02}} + \ketbra{S_{02}}{S_{11}}
  \right) - \Delta \ketbra{S_{02}}{ S_{02}},
\end{equation}
where $\Delta$ is the interdot energy detuning (see Fig.~\ref{fig:PSB}) and $t_\text{c}$ is the interdot tunnel splitting (see Fig.\ 2a of
the main article).  The interaction between the local magnetic fields and the
electron spins in the two dots is described by
\begin{equation}
  H_\text{B} = g \mu_\text{B} \left[ \left( \Bext + \Bml \right) \cdot \mathbf{S}^\text{L} + \left( \Bext + \Bmr \right) \cdot \mathbf{S}^\text{R} \right],
  \label{eq:Hb}
\end{equation}
where $\Bext$ is the external magnetic field, $\Bm^\text{L,R}$ the local
magnetic field of the nanomagnet, $\mathbf{S}^\text{L,R}$ the local electron
spin operator, $g$ the electron g-factor, and $\mB$ the Bohr magneton.

The wave function of an electron confined in a lateral GaAs-dot overlaps with
$\sim 10^6$ nuclei. The hyperfine interaction between the electron spin and
the nuclear spins is dominated by the contact term $H^\text{L,R}_\text{con} =
\sum_k A^\text{L,R}_k \mathbf{I}^\text{L,R}_k \cdot \mathbf{S}^\text{L,R}$, where
$\mathbf{I}^\text{L,R}_k$ is the $k$th nuclear spin operator and $A^\text{L,R}_k$ the
hyperfine coupling constants. $A^\text{L,R}_k$ is proportional to the overlap between the
wavefunctions of the $k$th nucleus in left/right dot and the electron and varies by isotope
type. It is common to define an average $A$ reflecting the natural abundance
of isotope type and the average overlap with the electron wavefunction. In
this approximation, the contact Hamiltonian is $H^\text{L,R}_\text{con} = A
\mathbf{I}^\text{L,R} \cdot \mathbf{S}^\text{L,R}$, where
$\mathbf{I}^\text{L,R}$ is the average nuclear spin (ensemble) operator and $A
= \SI{85}{\micro\electronvolt}$ in GaAs\cite{R2}\@. Electrons in the
left/right dot couple to different sets of nuclei, and we can write
\begin{equation}
  H_\text{hf} = H^\text{L}_\text{con} + H^\text{R}_\text{con}  = A \left( \mathbf{I}^\text{L} \cdot \mathbf{S}^\text{L} + \mathbf{I}^\text{R} \cdot \mathbf{S}^\text{R} \right) = A \sum_{i = \text{L,R}} \left( I_z^i S_z^i + \frac{I_+^i S_-^i + I_-^i S_+^i}{2} \right),
  \label{eq:raw Hhf}
\end{equation}
where $S^\text{L,R}_z$ and $I^\text{L,R}_z$ are the corresponding $z$-projection
operators; and $S^\text{L,R}_\pm = S^\text{L,R}_x \pm i S^\text{L,R}_y$ and
$I^\text{L,R}_\pm = I^\text{L,R}_x \pm i I^\text{L,R}_y$ are the spin raising
and lowering operators.
In a semiclassical approximation $\mathbf{I}^\text{L,R}$ can be replaced by
the effective nuclear magnetic (Overhauser) field\cite{R3} $\Bn^\text{L,R} = A \avg{
  \mathbf{I}^\text{L,R} } / \left( g \mB \right)$, where
$\avg{\ldots}$ denotes the expectation value, and
$\avg{I}_\text{max} = 3/2$ in GaAs. In ESR experiments (not shown), we
measured $g \simeq -0.36$ in our system. This predicts, for fully polarized
nuclear spins ($P = 1$), an Overhauser field magnitude of
$B^\text{max}_\text{nuc} = A\avg{I}_\text{max}/ \left( \abs{g} \mB \right)
\simeq \SI{6.1}{\tesla}$. The semiclassical version of $H_\text{hf}$ has the
same form as $H_\text{B}$ (see Eq.~\eqref{eq:Hb}), and we can summarize
\begin{equation}
  H_\text{B} + H_\text{hf} = g \mB \left(\Bl \cdot \mathbf{S}^\text{L} + \Br \cdot \mathbf{S}^\text{R} \right),
  \label{eq:semiclassical Hhf1}
\end{equation}
where $\mathbf{B}^\text{L,R} = \Bext + \Bm^\text{L,R} + \Bn^\text{L,R}$ is the
total effective magnetic field acting on an electron in the left and right
dot, respectively.

In analogy to $\Bs = \left( \Bl + \Br \right)/2$ and $\Bd = \left( \Bl - \Br
\right)/2$, we define the symmetric and antisymmetric spin operators $\Ss =
\left( \Sl + \Sr \right)/2$ and $\Sd = \left( \Sl - \Sr \right)/2$. We then
use $\overline{B}_\pm = \overline{B}_x \pm i \overline{B}_y$, $\Delta B_\pm =
\Delta B_x \pm i \Delta B_y$, $\Ss_\pm = \Ss_x \pm i\Ss_y$ and $\Sd_\pm =
\Sd_x \pm i \Sd_y$, defined akin to the spin raising and lowering operators in
equation~(\ref{eq:raw Hhf}), to write equation (\ref{eq:semiclassical Hhf1})
analogous to the right hand side of equation (\ref{eq:raw Hhf}):
\begin{equation}
  H_\text{B} + H_\text{hf} = g \mB \left( 2\overline B_z\Ss_z + 2\Del B_z \Sd_z + \Del B_+\Sd_- + \Del B_-\Sd_+ + \overline{B}_+ \overline{S}_- + \overline{B}_- \overline{S}_+ \right).
  \label{eq:semiclassical Hhf2}
\end{equation}
With the quantization axis, $\hat{z}$, defined parallel to $\Bext$, the matrix
representation of the (semiclassical) total Hamiltonian in the basis spanned
by the diabatic singlet and triplet states \{\Tp, \Tzero, \Tm, \Slr, \Srr \}
is

\parbox{.9\linewidth}{
\begin{center}
 \raisebox{-1.2em}{$\displaystyle H = \mu^\star\;$}
\begin{blockarray}{ccccc@{\hspace*{20pt}}|@{\hspace*{5pt}}c}
\hspace{5ex}\Tp\hspace{5ex} & \hspace{5ex}\Tzero\hspace{5ex} & \hspace{5ex}\Tm\hspace{5ex} & \hspace{5ex}\Slr\hspace{5ex} & \hspace{5ex}\Srr\hspace{3ex} &\\\cline{1-6}
&&&&&\\
 \begin{block}{(ccccc)@{\hspace*{20pt}}|@{\hspace*{5pt}}l}
 $\displaystyle \sqrt{2} \ \overline{B}_z$ & $\overline{B}_-$ & $0$ & $-\Delta B_-$ & $0$ & $\Tp = \ket{\uu}$ \\[0.7em]
 $\overline{B}_+$ & $0$ & $\overline{B}_-$ & $\sqrt{2} \Delta B_z$ & $0$ & $\Tzero = \left( \ket{\ud} + \ket{\du} \right)/\sqrt{2}$ \\[0.7em]
 $0$ & $\overline{B}_+$ & $-\sqrt{2} \ \overline{B}_z$ & $ \Delta B_+$ & $0$ & $\Tm  = \ket{\dd}$ \\[0.7em]
 $-\Delta B_+$ & $\sqrt{2} \Delta B_z$ & $\Delta B_-$ & $0$ & $\tc^\star/2$ & $\Slr = \left( \ket{\ud} - \ket{\du} \right)/\sqrt{2}$ \\[0.7em]
 $0$ & $0$ & $0$ & $\tc^\star/2$ & $-\Del^\star$ & $\Srr = \ket{0,\ud}$ \\[0.3em]
 \end{block}
\end{blockarray}
\end{center}} \hfill\parbox{0pt}{
\begin{equation}
  \label{eq:semiclassical Hhf matrix}
\end{equation}}

\noindent
where $\mu^\star=g\mB/\sqrt2$, $\tc^\star=\tc/\mu^\star$ and
$\Del^\star=\Del/\mu^\star$.
The matrix representation (\ref{eq:semiclassical Hhf matrix}) illustrates that
the $x$- and $y$-components of \Bd\ mix $T_\pm$ with $S_{11}$, while the
$z$-component of \Bd\ mixes \Tzero\ (which has no spin component along the $z$-axis)
with $S_{11}$.
Instead, $\overline{B}_z$ leads to the Zeeman splitting of the $T_\pm$
states. Note that the off-diagonal terms $\overline B_\pm$, which mix $T_\pm$
with $T_0$, vanish if the quantization axis is chosen parallel to \Bs.

\section{IV.	Perturbation calculation of DNSP}\label{sec:perturbation}

In the main article, we use a simplified approximation that only considers the
hyperfine contribution from $T_+$. In this section, we calculate the
hyperfine-induced decay of all (1\,1) states using perturbation theory
(Fermi's golden rule) similar to reference \cite{R4} in which, however,
the effects of a nanomagnet were not included. Here we show that the
simplified model produces the pertinent features of the perturbation theory,
justifying the approximation used in the main article.

We start by writing equation~\eqref{eq:H} as $H = H_0 + H^+_\text{ff} +
H^-_\text{ff}$, where
\begin{align}\label{eq:A:Ham}
  H_0 &= H_\text{el} + H_\text{B} + 2 g \mB \left( \overline{B}^z_\text{nuc} \overline{S}_z + \Delta B^z_\text{nuc} \Delta S_z \right)\,, \\
  H^\pm_\text{ff} &= g \mB \left( \overline B_\text{nuc}^\pm \Ss_\mp + \Del
    B_\text{nuc}^\pm \Sd_\mp \right)\,,
\end{align}
are the bare Hamiltonian and hyperfine flip-flop Hamiltonians,
respectively. We treat $H^+_\text{ff}+H^-_\text{ff}$ as a perturbation of
$H_0$.
Diagonalization of $H_0$ provides the unperturbed eigenvalues, $E_n$, of the
$n$th eigenstate, $\ket{n}$. We account for coupling to the leads by a simple
master equation with four Lindblad operators (eliminating the intermediate
stage in the sequential tunneling process $(0\,2)\to(0\,1)\to(1\,1)$) and
assuming that the four (1\,1) states $\{T_+,T_0,T_-,S_{11}\}$ are populated
with equal rate:
  \begin{equation}
    \label{eq:A:MEq}
    \frac{d}{dt}\rho =
    \frac{1}{i\hbar}[H_0,\rho]+\frac{\Gamma_R}{4}\sum_{x\in\{T_+,T_0,T_-,S_{11}\}}\left(\ketbra{x}{S_{02}}\rho\ketbra{S_{02}}{x}-\frac{1}{2}\left(\proj{S_{02}}\rho+\rho\proj{S_{02}}\right)^{\phantom{X}}\!\!\!\!\right).
  \end{equation}
We approximate the dynamics by a rate equation for the populations
$\boldsymbol{\rho}=(\rho_{11},\rho_{22},\rho_{33},\rho_{44},\rho_{55})$ in
the five energy eigenstates.
\begin{equation}
  \label{eq:A:RateEq0}
\frac{d\boldsymbol{\rho}}{dt} = G^{(0)} \boldsymbol{\rho}.
\end{equation}
The transition matrix $G^{(0)}=(G^{(0)}_{ij})_{ij}$, describes decay of the
level $n$ with a rate $\Gamma_n^0=\Gamma_Rs_n$, determined by
$s_n=|\braket{S_{02}}{n}|^2$, the overlap of $\ket{n}$ with the localized
singlet state. Since only (1\,1) states are refilled, the rate with which
$\ket{m}$ is populated is proportional to $(1-s_m)$: Hence the matrix elements
of $G^{(0)}$ are
\begin{align}
  \label{eq:Gamma}
G^{(0)}_{nm} &=\phantom{-}\frac{\Gamma_R}{4}(1-s_n)s_m\,\,\,(n\not=m),\\
G^{(0)}_{nn} &= -\frac{\Gamma_R}{4}s_n\left( 3+s_n\right).
\end{align}
The width of the level $\ket{n}$ is given by $\hbar\Gamma^0_n=\hbar\Gamma_Rs_n$.

We include the hyperfine flip-flop processes using Fermi's golden rule (assuming a constant density of states over the range of energies of the (1\,1) states) to determine the flip-flop rate from an initial state, $\ket{n}$,
to the final state, $\ket{f}$ as
\begin{align*}
  \Gamma^\pm_{n\to f} & = \dfrac{2\pi}{\hbar} \abs{\matrixel{f}{H^\pm_\text{ff}}{n}}^2  \dfrac{1}{2\pi}\dfrac{\hbar \left( \Gamma^0_f + \Gamma^0_n \right)}{\left( E_n - E_f \right)^2 + \left( \hbar \frac{ \Gamma^0_f + \Gamma^0_n}{2} \right)^2} \frac{1 \mp P}{2} \\
  & = \dfrac{\abs{\matrixel{f}{H^\pm_\text{ff}}{n}}^2 \left(
      \Gamma^0_f + \Gamma^0_n \right)}{ \left( E_n - E_f \right)^2 + \left(
      \hbar \frac{ \Gamma^0_f + \Gamma^0_n}{2} \right)^2} \frac{1 \mp P}{2},
\end{align*}
where the factor $\left(1 \mp P \right) / 2$ expresses the influence of
polarization on the nuclear-spin flip rates. The total escape rate from
$\ket{n}$ is then
\begin{equation}
  \Gamma_n = \Gamma^0_n + \sum_{f\not=n}(\Gamma^+_{n\to f} + \Gamma^-_{n\to f}).
\end{equation}
We can neglect cotunneling in our sample, since its contribution to the
leakage current is negligible compared to that of the nanomagnet.
The full rate equation is now given by
\begin{equation}\label{eq:A:FullRateEq}
  \frac{d\boldsymbol{\rho}}{dt} = G \boldsymbol{\rho},
\end{equation}
with $G_{nm}=G^{(0)}_{nm}+\Gamma^+_{n\to m} + \Gamma^-_{n\to m} (n\not=m)$ and
$G_{nn} = G^{(0)}_{nn}-\sum_{m\not=n}(\Gamma^+_{n\to m} + \Gamma^-_{n\to m})$.

We determine the leakage current and the DNSP rates by numerically solving for
$\dot{\boldsymbol{\rho}} = 0$, obtaining the steady state populations
$\rho^\text{ss}_n$.
The magnitude of the total current is given by
\begin{equation*}
  I = I^0 + I^+ + I^- = e \sum_n{ \rho^\text{ss}_n \Gamma_n},
\end{equation*}
where $e$ is the magnitude of the electron charge. $I^0$ is the non-polarizing
current induced by the nanomagnet, while $I^\pm$ are the hyperfine-generated
currents that polarize in opposite directions. $I^\pm$ are expressed as
\begin{equation*}
  I^\pm = e \sum_n{ \rho_n \Gamma^\pm_n}.
\end{equation*}
To convert these currents into nuclear polarization rates, we write
\begin{equation*}
  \Gamma^\pm_\text{pol} = \frac{2I^\pm}{eN},
\end{equation*}
where we have normalized by $e$ and included that the polarization, $P = \left(
  N^\uparrow - N^\downarrow \right)/N$, changes by $2/N$ per nuclear spin flip
for $N$ nuclei in total.  The overall polarization rate equation is
\begin{equation}
  \dot{P} = \Gamma^+_\text{pol} - \Gamma^-_\text{pol} - P \Gamma_\text{rel}.
  \label{eq:full pdot}
\end{equation}
In the main article, we neglect $\Gamma^-_\text{pol}$ and approximate
\begin{equation}
  \dot{P} \simeq \Gamma^+_\text{pol} - P \Gamma_\text{rel} \simeq \alpha I \left( 1 - P \right) - P \Gamma_\text{rel},
  \label{eq:appr pdot}
\end{equation}
where $\alpha$ is taken as a constant ($\sim \SI{35}{\per\nano\ampere\per\second}$).

\begin{figure}[h]
  \includegraphics{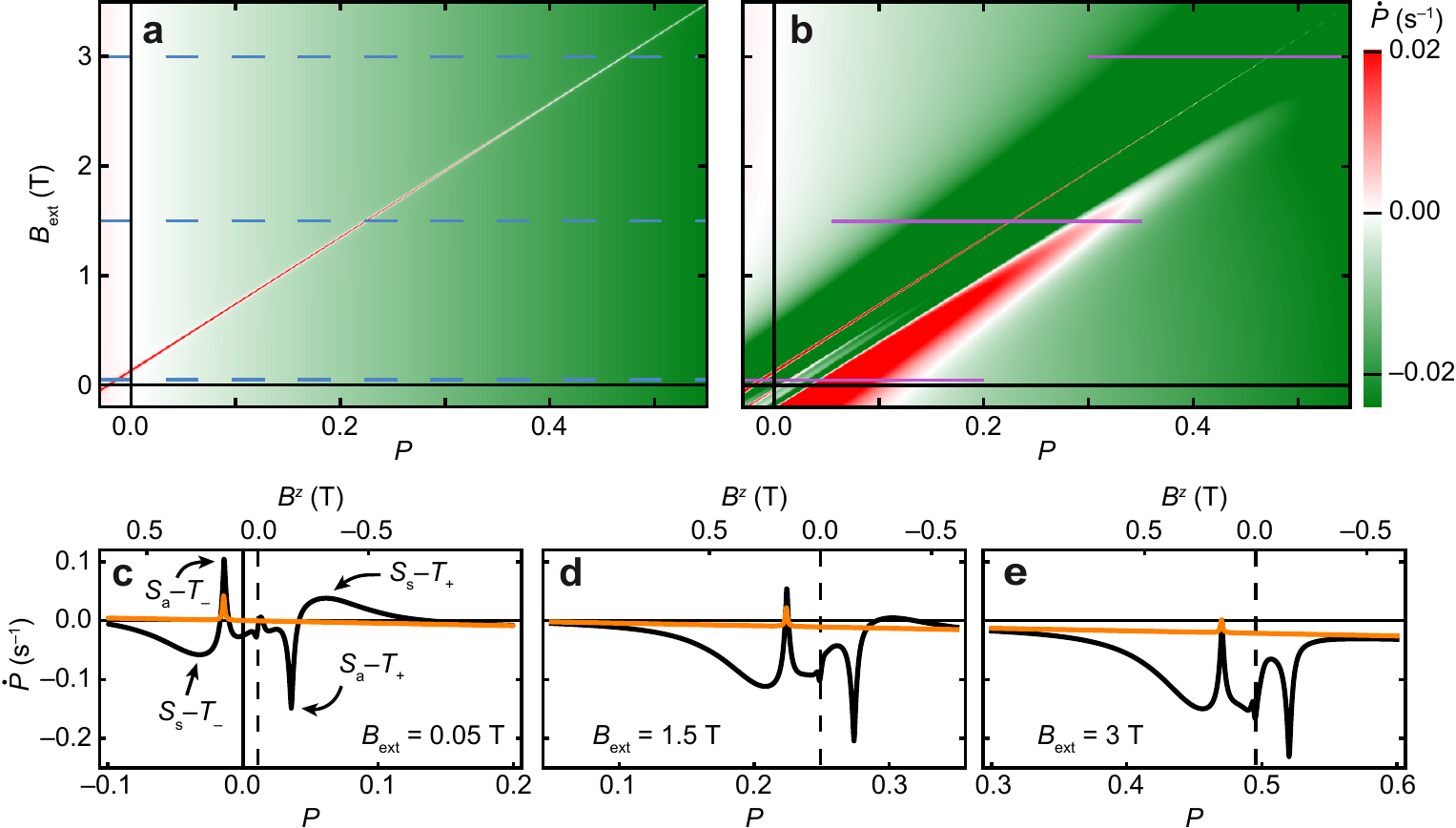}
  \caption{\textbf{Nuclear polarization rate a)} The nuclear polarization
    rate, $\dot{P}$, calculated analytically using equation~(1) of the main
    article. The dashed horizontal lines correspond to the constant
    $B_\text{ext}$ slices shown in Fig.~3b of the main article.
    \textbf{b)} $\dot{P}$ calculated numerically using equation~\eqref{eq:full
      pdot} with $\Gamma_\text{R} = \SI{4.5}{\giga\hertz}$ and $N = 10^6$
    nuclei. A random Gaussian distribution was used for $B^x_\text{nuc}$ and
    $B^y_\text{nuc}$ as in reference \cite{R5}.
    The scale bar applies to both \textbf{a} and \textbf{b}. Both calculations
    were performed using $t_c = \SI{12}{\micro\electronvolt}$, $\Delta =
    \SI{8}{\micro\electronvolt}$, $\Grel = \SI{0.04}{\per\second}$, and the
    magnetic field distribution of the nanomagnet (Fig.\ 1c, main article).
    \textbf{c-d)} $\dot P$ as a function of $P$ at constant \Bext\ (lines in
    \textbf{a},\textbf{b}) comparing the rate equation model [orange,
    equation~(1) of the main article] with the perturbation calculation
    [black, equation~\eqref{eq:full pdot}]. The ranges in $P$ of these slices
    are indicated by the magenta lines in \textbf{b}. The polarization rate
    extrema associated with the four singlet-triplet resonances are labeled in
    \textbf{c}. The top axis indicates the $z$-component of the total
    effective magnetic field, $B^z = B_\text{ext} + \overline{B}^z_\text{nm} -
    P B_\text{nuc}^\text{max}$. In addition to the extrema associated with the
    singlet-triplet resonances small polarization features appear near $B^z =
    0$ (dashed vertical lines) where triplets become degenerate.}
  \label{fig:flipflop}
\end{figure}

To compare the above theoretical model with the simplified version used in
the main text, Fig.~\ref{fig:flipflop} shows $\dot{P}$ versus $P$ and
$B_\text{ext}$ predicted using the two models represented by equations
(\ref{eq:full pdot}) and (\ref{eq:appr pdot}). The simple model, shown in
Fig.~\ref{fig:flipflop}a, predicts a narrow region of positive $\dot{P}$
running diagonally through the $P$--$B_\text{ext}$ plane. The perturbation
theory calculation, shown in Fig.~\ref{fig:flipflop}b, predicts this same
feature, which is associated with the $\Sa$--$T_-$ resonance, as well as more
complicated behavior associated with other singlet--triplet resonances (see
Fig.~\ref{fig:flipflop}c). Both models predict that the $\Sa$--$T_-$ resonance
is the only resonance which creates a stable fixed point at large $P$ and
$B_\text{ext}$. Figs.~\ref{fig:flipflop}c-e demonstrate the good agreement
between the simplified model (orange lines) and the perturbation calculation
(black lines) at the $\Sa$--$T_-$ resonance. (The orange lines here are the
same as the lines in Fig.~4a of the main article.) Compared to the other two
resonances, $\dot{P}$ at the $\Sa$--$T_\pm$ resonances is very steep and
provides strong feedback of the nuclear spins toward the A-FP, a distinct
advantage for DNSP.  From Fig.~\ref{fig:flipflop}, it is evident that
equations~(1--4) of the main article are sufficient to model the nuclear
polarization associated with the $\Sa$--$T_-$ resonance.

The leftmost and rightmost $\dot{P}$ extrema in Figs.~\ref{fig:flipflop}c-d
are generated by the $\Sp$--$T_\pm$ resonances, while the small feature
between the $\Sa$--$T_\pm$ resonances corresponds to the crossing of the
triplet levels at $B^z = B_\text{ext} + \overline{B}^z_\text{nm} - P
B_\text{nuc}^\text{max} \approx 0$.
Note that the nuclear field components $\bar{B}_\text{nuc}^{x,y}, \Delta
B^{x,y,z}_\text{nuc}$ play an important role for our treatment, determining
directly the strength of the hyperfine flip-flop rates. For the DNSP rates
depicted in Fig.~\ref{fig:flipflop} we have averaged calculations for
different values of the nuclear field fluctuations chosen according to a
Gaussian distribution with zero mean and standard deviation of $\sim
\SI{3}{\milli\tesla}$.

\newpage
\section{V.	Hyperfine- and Non-hyperfine--induced Leakage
  Current}\label{sec:current}

\begin{figure}[b]
  \includegraphics[width = 0.5\textwidth]{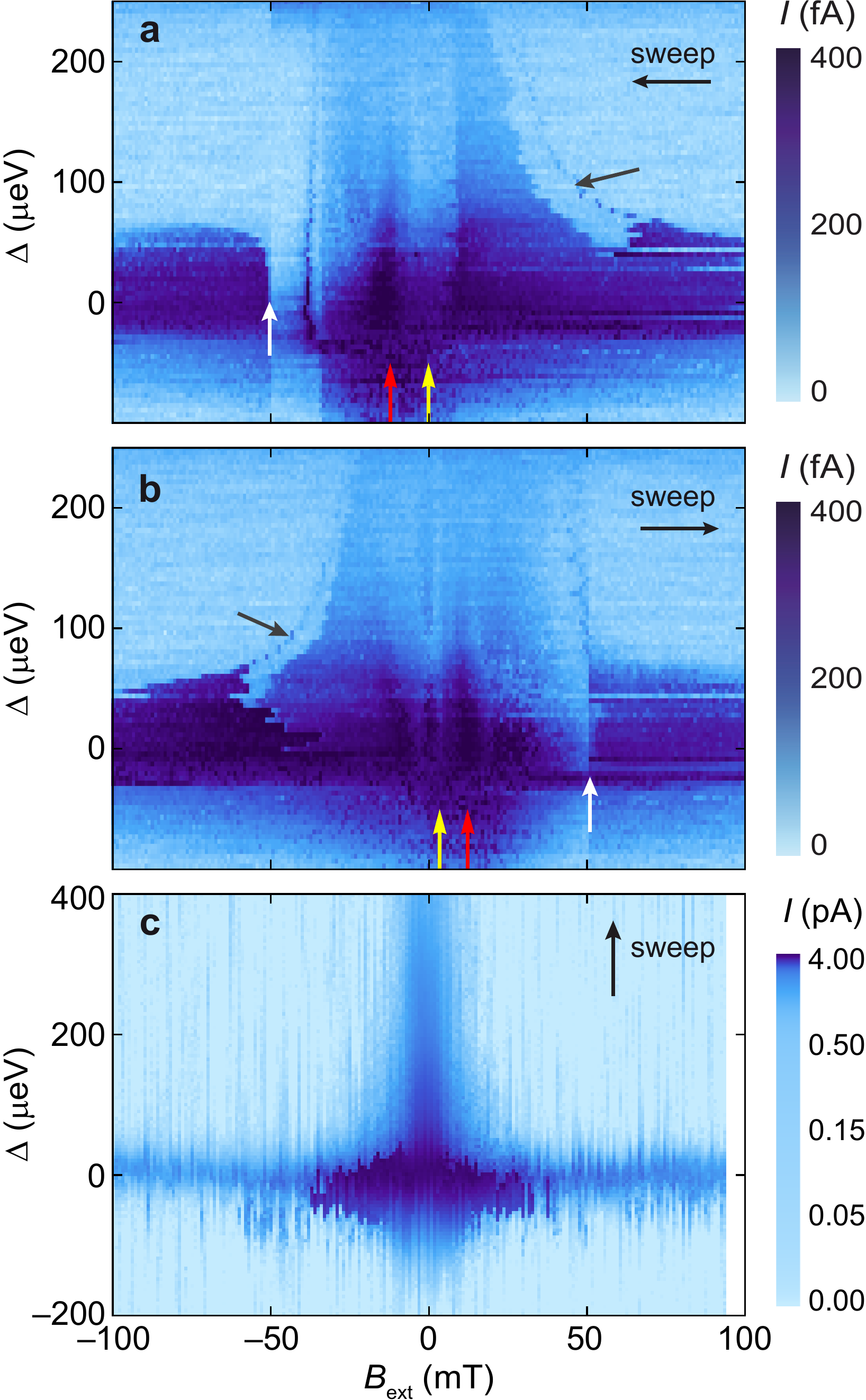}
\caption{
\textbf{Spin-blockaded leakage current.}
\textbf{a)} The dc leakage current, $I$, as a function of $B_\text{ext}$ (swept from positive to negative) and \Del\ (stepped from positive to negative) ($\tc \sim \SI{15}{\micro\electronvolt}$.
\textbf{b)} Same as in \textbf{a} but for the opposite sweep direction of \Bext\ (from negative to positive). Arrows in \textbf{a} and \textbf{b} mark specific features explained in the main text.
\textbf{c)} $I$ as a function of $B_\text{ext}$ and \Del\ measured using a sample with an identical gate layout as the sample used in the main article, but without the on-chip nanomagnet. Overall, maximum $I$ values are one order of magnitude larger than in \textbf{a} and \textbf{b} owing to stronger source/drain coupling with this particular gate tunning, but the region of enhanced $I$ is much smaller. The magnetic field was stepped from right to left and the energy detuning was swept from bottom to top. The perpendicular sweep direction compared to \textbf{a} and \textbf{b} does not affect the main features of this measurement. It does however cause a noisy background which is typical for this sweep direction and is caused by charge noise triggered by sweeping gate voltages.
\label{fig:B vs Delta comparison}
}
\end{figure}

The leakage current through the DQD shown in Fig.\ 2b of the main article
contains a number of features which can be traced back to the inhomogeneous
field produced by the single domain nanomagnet. To illustrate this, in
Fig.\ \ref{fig:B vs Delta comparison} we show the PSB leakage current through
two different DQD devices with identical gate layout. The data in Figs.\
\ref{fig:B vs Delta comparison}\textbf{a} and \ref{fig:B vs Delta
  comparison}\textbf{b} show measurements for opposite sweep directions
acquired on the sample also presented in the main article but for a larger
interdot tunnel coupling $\tc \sim \SI{15}{\micro\electronvolt}$. These data
are more richly featured compared to those in Fig.\ \ref{fig:B vs Delta
  comparison}c in which no nanomagnet was present ($\tc \sim
\SI{1}{\micro\electronvolt}$).

In Fig.~\ref{fig:B vs Delta comparison}c, the measured current is
approximately symmetric with respect to the $B_\text{ext} = 0$ axis, and the
main features are increased current along the $\Delta = 0$ and $B_\text{ext} =
0$ axes (for $\Delta \gtrsim 0$, that is, outside of Coulomb blockade) and a
global maximum at $\Delta = B_\text{ext} \simeq 0$. Similar data have already
been published and discussed in detail in reference \cite{R6}. In
short, current is created by $\tc$ in combination with the hyperfine
interaction, which mixes triplet and singlet states strongly when $\Del \simeq
0$ or $B_\text{ext} \simeq 0$. The width of the current maximum along the
$B_\text{ext} = 0$ axis is determined by the standard deviation of the
fluctuating $\Bn$\cite{R6,R5}. The data in Figs.\ \ref{fig:B
  vs Delta comparison}a and \ref{fig:B vs Delta comparison}b illustrate that
the sizeable $\Bm$ adds complexity. The following list provides a short
explanation for each of the features specific for the sample with nanomagnet:
\begin{itemize}
\item The most obvious response to expect when sweeping $B_\text{ext}$ is
  hysteresis of the magnetization of the single domain nanomagnet. Because of
  its single-domain character, we expect the nanomagnet to switch polarization
  abruptly when $B_\text{ext}$ passes the coercive field. An abrupt switch in
  magnetic field, in turn, should cause an equally abrupt change in the
  leakage current signal. Such features are indeed observed in
  Figs.~\ref{fig:B vs Delta comparison}a,b at $B_\text{ext} \simeq \mp
  \SI{52}{\milli\tesla}$, respectively (see white arrows), and are also seen
  in Fig.~2b of the main article.
\item In the presence of $\Bm$, the eigenenergies of the $T_\pm$--like states
  are never zero. However, the relevant magnetic field, $\abs{\Bl + \Br}$ can
  be minimized by $B_\text{ext}$, and at this minimum, the $T_\pm$ are most
  degenerate and a local maximum of the leakage current is expected. For our
  $\Bm$ values, $\abs{\Bl + \Br}$ is minimized at $B_\text{ext} \simeq
  \pm\SI{12}{\milli\tesla}$, depending on the polarization of the
  nanomagnet. These fields are indicated in Figs.~\ref{fig:B vs Delta
    comparison}a,b with red arrows and faithfully identify the current maxima.
\item The observation of distinct local current maxima at the \Sm--$T_\pm$
  resonance (black arrows in Figs.\ \ref{fig:B vs Delta comparison}a and
  \ref{fig:B vs Delta comparison}b) and in Fig.~2b of the main article) is
  unique to samples containing a single domain nanomagnet. The sharpness of
  these peaks can only be explained by taking into account the hyperfine
  induced dynamics of the nuclear spins. The actual position of the
  \Sm--$T_\pm$ resonance is shifted towards larger $\abs{B_\text{ext}}$
  compared to its prediction (see Fig.\ 2b of the main article). In the main
  article we explain this shift by taking into account the hyperfine induced
  DNSP.
\item $\Bm$ mixes triplet and singlet states weakening the spin blockade and
  allowing leakage current to flow. However, $B_\text{ext}$ tunes this
  mixing. In fact, the condition $\Bd \parallel \Bs$ defines a local minimum
  of the singlet mixture with the $T_\pm$ states~\cite{R5}. This can
  be readily seen from the Hamiltonian in equation (\ref{eq:semiclassical Hhf
    matrix}) if the quantization axis is chosen parallel to \Bs. For our
  system, this condition is satisfied when $B_\text{ext} \simeq
  \pm\SI{8}{\milli\tesla}$. We actually observe current minima at slightly
  shifted values (see yellow arrows in Fig.~\ref{fig:B vs Delta
    comparison}a,b) owing to the complex DNSP that occurs while sweeping
  $B_\text{ext}$ (see Fig.~\ref{fig:flipflop}).
\end{itemize}

The current at finite $B_\text{ext}$ and small \Del\ in Figs.\ \ref{fig:B vs Delta comparison}a and \ref{fig:B vs Delta comparison}b is characterized by
strong switching noise and dragging effects which has also been observed in
samples without on-chip magnet\cite{R6,R7}. We forgo a
detailed discussion of these effects, which can be explained in terms of the
hyperfine dynamics in the presence of more than one stable fixed point at
$\Del \sim 0$\cite{R4}.

\clearpage \newpage
\section{VI.	Nuclear Polarization Build-up}\label{sec:buildup}

\begin{figure}[h]
\includegraphics[width = 0.7\textwidth]{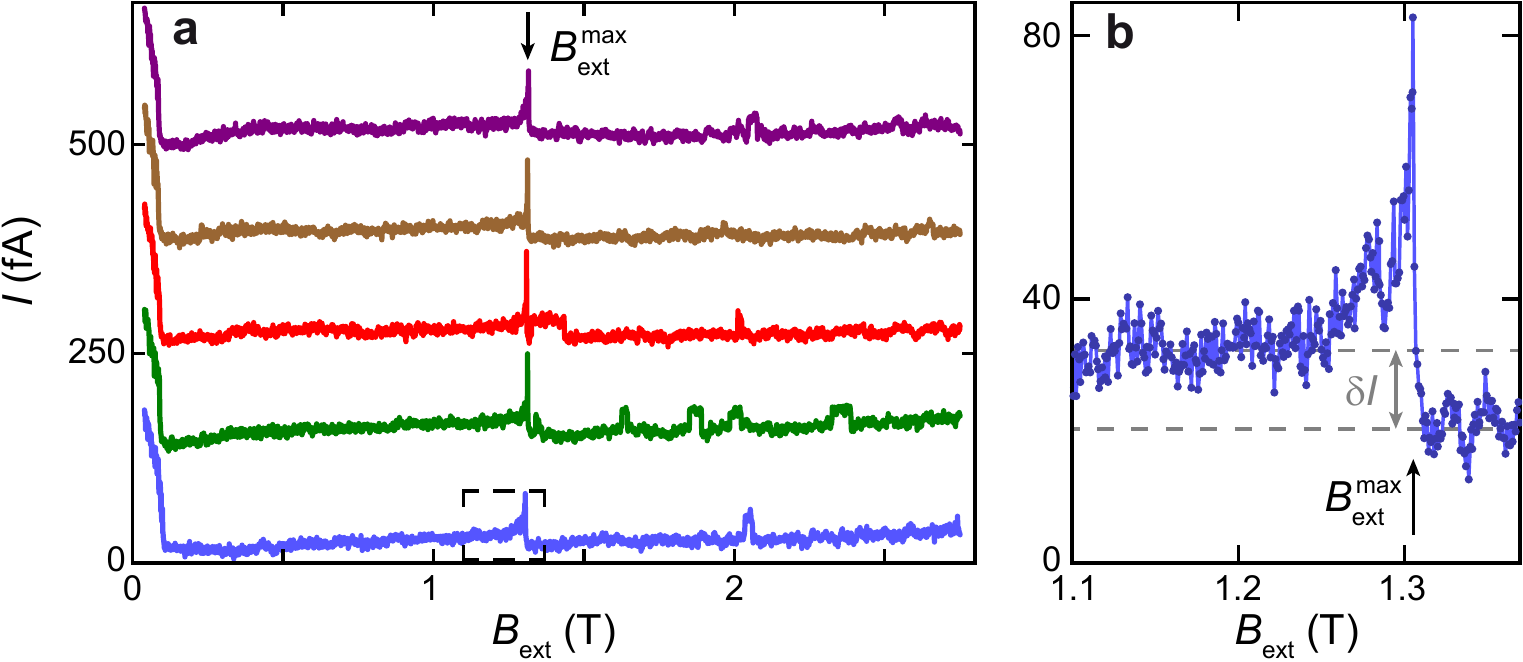}
\caption{
\textbf{Polarization sweep repeatability a)}
Five polarization sweeps measured at a rate $\dot{B}_\text{ext} = \SI[per-mode=symbol]{35}{\milli\tesla\per\minute}$ and $\Delta = \SI{150}{\micro\electronvolt}$. The traces are offset by $N \times \SI{120}{\femto\ampere}; N = 0,1,2,3,4$ for clarity. The current bistabilities observed beyond $B^\text{max}_\text{ext}$ are consistent with DNSP\cite{R6,R4}. \textbf{b)} The details of a current trace near $B^\text{max}_\text{ext}$ taken from within the boxed region of \textbf{a} demonstrate a sharp resonance and clear change in current, $\delta I \simeq \SI{11}{\femto\ampere}$, before and after sweeping through $B^\text{max}_\text{ext}$.
\label{fig:repeated sweeps}
}
\end{figure}

\begin{figure}[h]
\includegraphics[width = 0.8\textwidth]{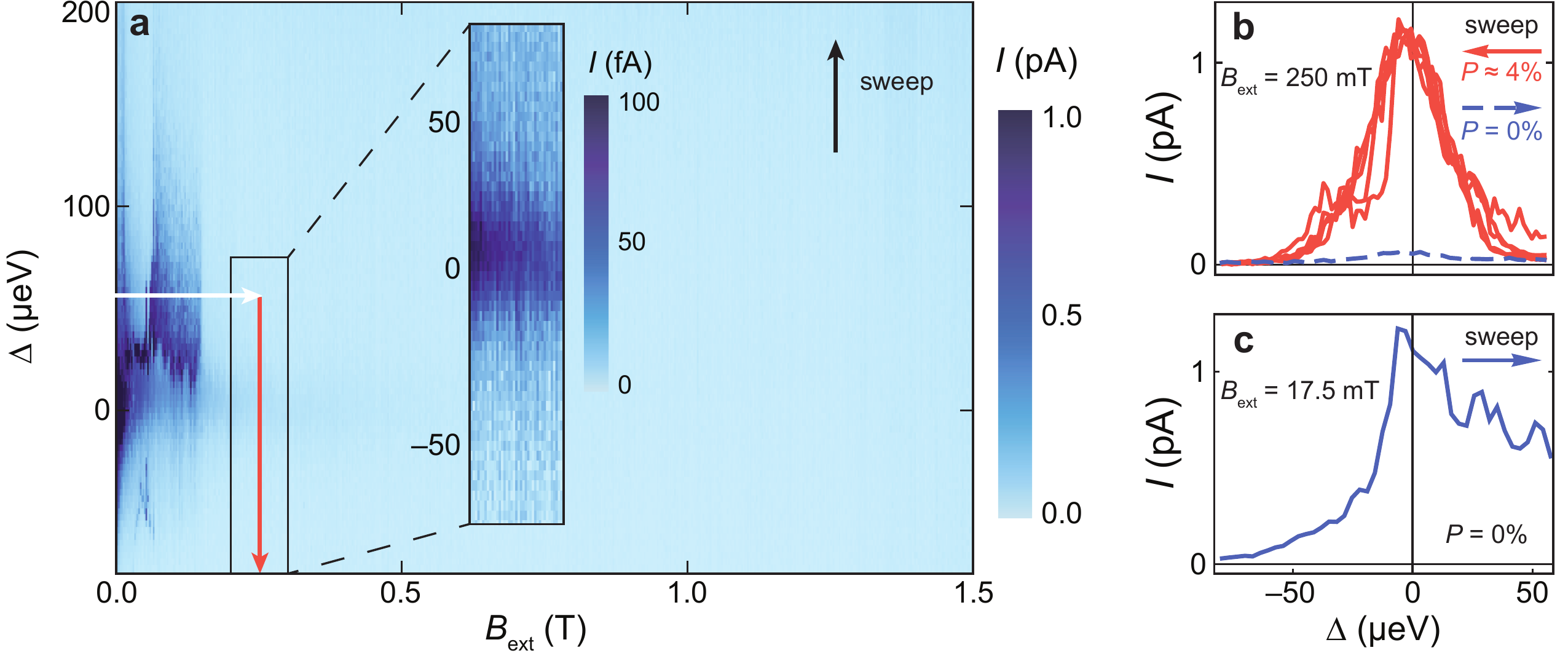}
\caption{
\textbf{PSB leakage current at large $B_\text{ext}$ a)} Leakage current, $I$, as a function of \Del\ and a large range of $B_\text{ext}$. The magnetic field has been stepped and \Del\ has been swept from negative to positive to minimize DNSP effects. \textit{Inset}: Enlarged region centered at $B_\text{ext} = \SI{250}{\milli\tesla}$ showing the small ($\sim\SI{100}{\femto\ampere}$) leakage current when $P = \SI{0}{\percent}$.
\textbf{b)} The solid orange lines are $I$ measured while sweeping $\Delta$ multiple times through $\Delta = 0$ into Coulomb blockade along the orange arrow in \textbf{a}. These data were measured after ramping $B_\text{ext}$ from $B_\text{ext} \simeq 0$ to $B_\text{ext} \simeq \SI{250}{\milli\tesla}$ (along the white arrow in \textbf{a}) creating a polarization of $P \simeq \SI{4}{\percent}$. The dashed blue line is $I$ at $P = \SI{0}{\percent}$ extracted from \textbf{a} at $B_\text{ext} = \SI{250}{\milli\tesla}$ and is negligible compared to $I$ with polarization.
\textbf{c)} $I$ measured versus $\Delta$ starting with unpolarized nuclei extracted from \textbf{a} at $B_\text{ext} = \SI{17.5}{\milli\tesla}$.
\label{fig:extended B vs Delta}
}
\end{figure}

Here we present additional data to support our interpretation of the DNSP data. Fig.~\ref{fig:repeated sweeps} demonstrates that the polarization dragging in Fig.~4b of the main article is reproducible. All the main features in Fig.~\ref{fig:repeated sweeps}a are reproducible, especially the position of $B^\text{max}_\text{ext}$ where polarization is lost. The current measured at the beginning of each field sweep near $B_\text{ext} = 0$ is the typical leakage current that appears near $B_\text{ext} = 0$ (see, for example, Fig.~\ref{fig:extended B vs Delta}) and is extended somewhat because of DNSP. We interpret the sharp current maximum (labeled $B_\text{ext}^\text{max}$) as the point of maximum polarization (see main article). The $\delta I$ in Fig.~\ref{fig:repeated sweeps}b results from losing the stable polarization condition and related resonant current as the polarization decays and the system drifts away from resonance.

Fig.~\ref{fig:extended B vs Delta} shows a second technique for demonstrating the polarization created during fixed-point dragging measurements. In Fig.\ \ref{fig:extended B vs Delta}a, we show $I$ measured as a function of $B_\text{ext}$ and $\Delta$ over a much larger range of $B_\text{ext}$ than Fig.~2b of the main article. The current features at low $B_\text{ext}$ in \ref{fig:extended B vs Delta}\textbf{a} differ from those in Fig.~2b of the main article mostly because $\Delta$ was swept rather than $B_\text{ext}$.

The current traces in Fig.~\ref{fig:extended B vs Delta}b, measured at $B_\text{ext} = \SI{250}{\milli\tesla}$ after the nuclei have been polarized to $P \approx \SI{4}{\percent}$, are very similar to the trace in Fig.~\ref{fig:extended B vs Delta}c, which shows current measured at $B_\text{ext} = \SI{17.5}{\milli\tesla}$ and $P = \SI{0}{\percent}$. The quantitative similarity between the $P \approx \SI{4}{\percent}$ current traces in Fig.\ \ref{fig:extended B vs Delta}b and the trace in Fig.\ \ref{fig:extended B vs Delta}c allows us to conclude that $B^z_\text{nuc}$ compensates $B_\text{ext}$ ($B^z_\text{nuc} \simeq -B_\text{ext}$ for $B_\text{ext} \gg \abs{\Bm}$) thereby reducing the total effective field. In contrast to the polarized traces in Fig.\ \ref{fig:extended B vs Delta}b, which are almost symmetric with respect to $\Del = 0$, the $P = \SI{0}{\percent}$ curve in Fig.\ \ref{fig:extended B vs Delta}c is asymmetric and exhibits switching noise for $\Del > 0$. We attribute this behavior to small changes in nuclear spin polarization, while in Fig.~\ref{fig:extended B vs Delta}b, the polarization is stabilized at the adjustable fixed point (A-FP). The current traces in Fig.~\ref{fig:extended B vs Delta}b measured at $P \approx \SI{4}{\percent}$ are repeatable and much larger than the trace measured at $P = \SI{0}{\percent}$. This demonstrates that the polarization is finite and stable.

Current features, such as the local maxima in Fig.~2a of the main article and Fig.~\ref{fig:repeated sweeps} at $B_\text{ext} > 1\,$T, are not seen in Fig.\ \ref{fig:extended B vs Delta}a. These features are missing because $\Delta$ is swept, and significant polarizations are not obtained. Taken together with Figs.~2,3 of the main article, Figs.~\ref{fig:repeated sweeps} and \ref{fig:extended B vs Delta} demonstrate the ability of our system to generate large nuclear spin polarization---and detect it.

\newpage
\section{VII.	Data Fitting}\label{sec:fitting}

The fitting procedure of the data in article Figs.~2b and 3a involves solving numerically the nonlinear differential equation given by equation~(1) of the main article for a given set of parameters. This produces $P(t)$, which is then fed into equation~(2) to find $I(t)$. Our goal is not to reproduce the details of the measured $I(t)$ traces, but only the position of its maximum at the $\Sa$--$T_-$ resonance, that is, the position of the resonant current $\Imax$. Therefore, the final step is to calculate the position of $\Imax$ using the numerical $I(t)$. This procedure was repeated with different parameter sets until an agreement between theory and data was found.

The numerical fit needs the following parameters: $\tc$, $\gamma$, $\alpha$, $\Imax$, and $\Grel$ (see equations~(1--3) of the main article). $\Imax$ and the ratio $\alpha/\Grel = \SI{0.8}{\per\pico\ampere}$ (see main article Fig.~4d) were measured, thus reducing the overall number of fit parameters to three.

\begin{figure}[h]
  \includegraphics[width = 0.8\textwidth]{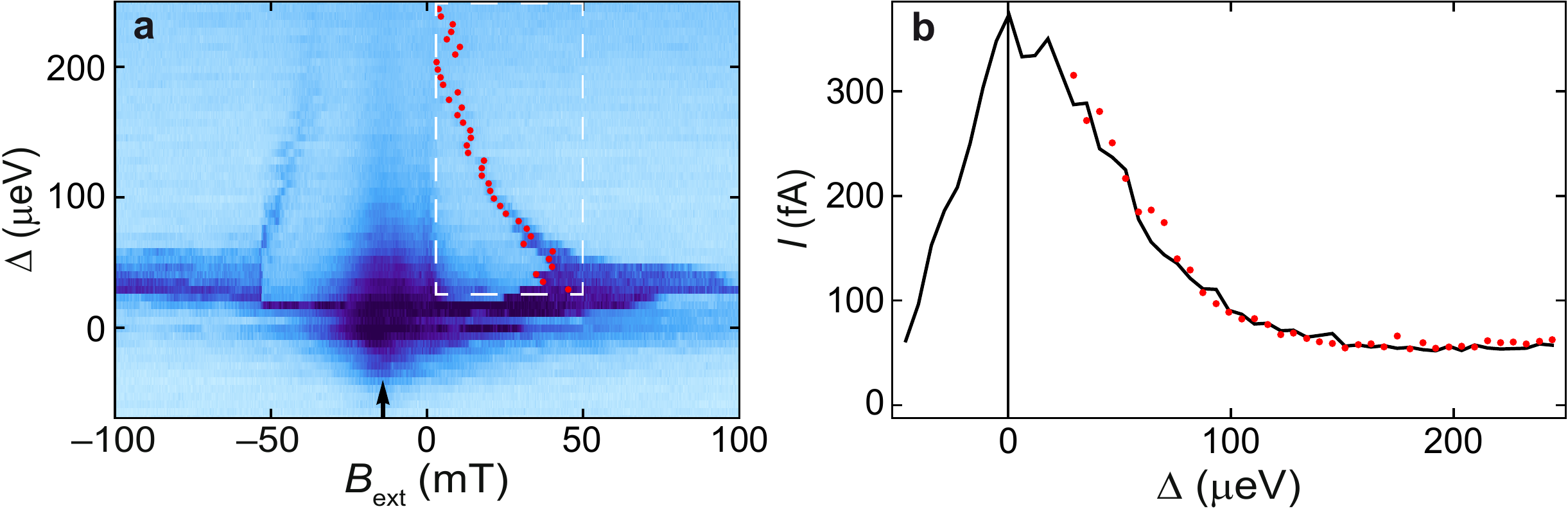}
\caption{
\textbf{Determination of $\Imax$. a)} Leakage current, $I$, measured versus $B_\text{ext}$ and $\Delta$ repeated from Fig.~2b of the main article. After searching within the boxed region, the {\color{red}$\CIRCLE$} indicate the $B_\text{ext}$-$\Delta$ position of $\Imax$ along the $\Sa$--$T_-$ resonance. \textbf{b}) Here {\color{red}$\CIRCLE$} are the $\Imax$ values from \textbf{a} plotted as a function of $\Delta$. In comparison, the black line is $I$ measured versus $\Delta$ along the symmetry axis (black arrow in \textbf{a}). As expected, $\Imax (\Delta)$ along the $\Sa$--$T_-$ resonance follows the general $I (\Delta)$ trend. A smoothed version of $I (\Delta)$ is used to create the smooth numerical fit in Fig.~2b of the main article.
\label{fig:current extraction}
}
\end{figure}

For the time-dependent data (see main article Figs.~3), $\Imax = \SI{100}{\femto\ampere}$ is the measured peak height. $\Imax$ values for the $\Delta$-dependent data (see main article Fig.~2b) are unique for each value of $\Delta$ because $I$ is $\Delta$ dependent. Fig.~\ref{fig:current extraction} details how $\Imax (\Delta)$ is extracted from the $I$ measured as a function of $B_\text{ext}$ and $\Delta$. The main result is that $\Imax (\Delta)$ is identical to $I(\Delta)$ measured near $B_\text{ext} = 0$, where spin blockade is lifted.

\begin{figure}[bh]
  \includegraphics[width = 0.65\textwidth]{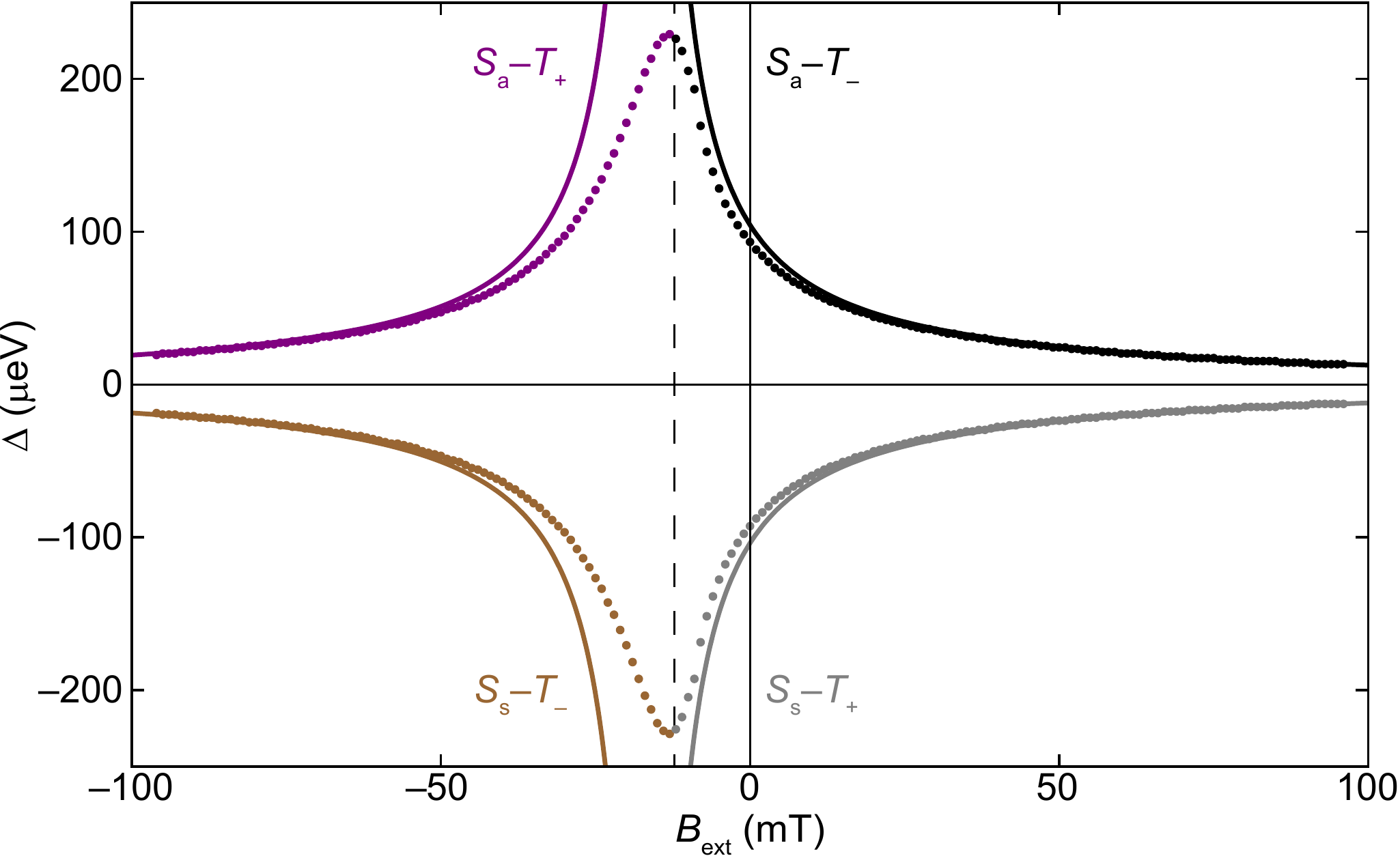}
\caption{
\textbf{Location of singlet--triplet resonances.}
The numerically calculated data points indicate the position in the $B_\text{ext}$--$\Delta$ plane where singlet and triplet states are resonant. The solid lines are approximated values calculated analytically. Each line is labeled to identify which states are resonant. In particular, the black line is the analytic approximation for the $\Sa$--$T_-$ resonance and was used to fit the DNSP data (see Fig.~2 of the main article). The exact and approximate values are in excellent agreement for $B_\text{ext} \geq 0$, where all DNSP measurements were performed. The upper branch of the numerical calculation is included in Fig.~2b of the main article. The vertical dashed line at $B_\text{ext} = \SI{-12}{\milli\tesla}$, which is the minimizing value of $\abs{\Bl +\Br}$, defines the symmetry axis where the triplets are most degenerate. Here $\tc = \SI{12}{\micro\electronvolt}$.
\label{fig:B0}
}
\end{figure}

When $\Bm$ and $\Bn$ are known, the $\Delta$--$B_\text{ext}$ position of the $\Sa$--$T_-$ resonance can be approximated analytically. This approximation is used in equation (4) of the main article and includes only $B^z_\text{nm}$ providing $\Ez^\pm \simeq \pm g \mB \left( B_\text{ext} - B^\text{max}_\text{nuc} P + \overline{B}^z_\text{nm}
\right)$. In Fig.~\ref{fig:B0}, we compare exact numerical results with the
analytical approximations for all four singlet--triplet resonances. The
analytical approximation for the $\Sa$--$T_-$ resonance is in excellent
agreement with the numerical calculation for $B_\text{ext} \geq 0$.


One set of fit parameters ($\tc = \SI{12}{\micro\electronvolt}$, $\Grel =
\SI{0.043}{\per\second}$, $\gamma =
\SI[separate-uncertainty]{10(1)}{\milli\tesla}$, and $\alpha =
\SI{35}{\per\nano\ampere\per\second}$) reproduces the data of two very
different experiments in Figs.~2b,3a of the main text. The data sets were
measured using identical gate voltages. For a slightly different system
tuning, only two parameters are expected to change, namely $\tc$ and $\alpha$,
because they reflect the various hyperfine and non-hyperfine system rates,
which are strongly gate dependent. $\gamma$ depends mostly on $\Bm$, which is
constant, while $\Grel$ should be independent of gate tuning because it is a
property of the nuclei. These expectations are supported in Fig.~\ref{fig:B vs
  Time} where $I$ versus $B_\text{ext}$ and time has been measured after
making the voltage of the top center gate more positive (see the gate design
in Fig.~1a of the main article). The ability to describe disparate sets of
data in different tuning regimes with either no change or only justifiable
adjustments to fit parameters demonstrates the validity of our model.
\begin{figure}[b]
  \includegraphics[width = 0.6\textwidth]{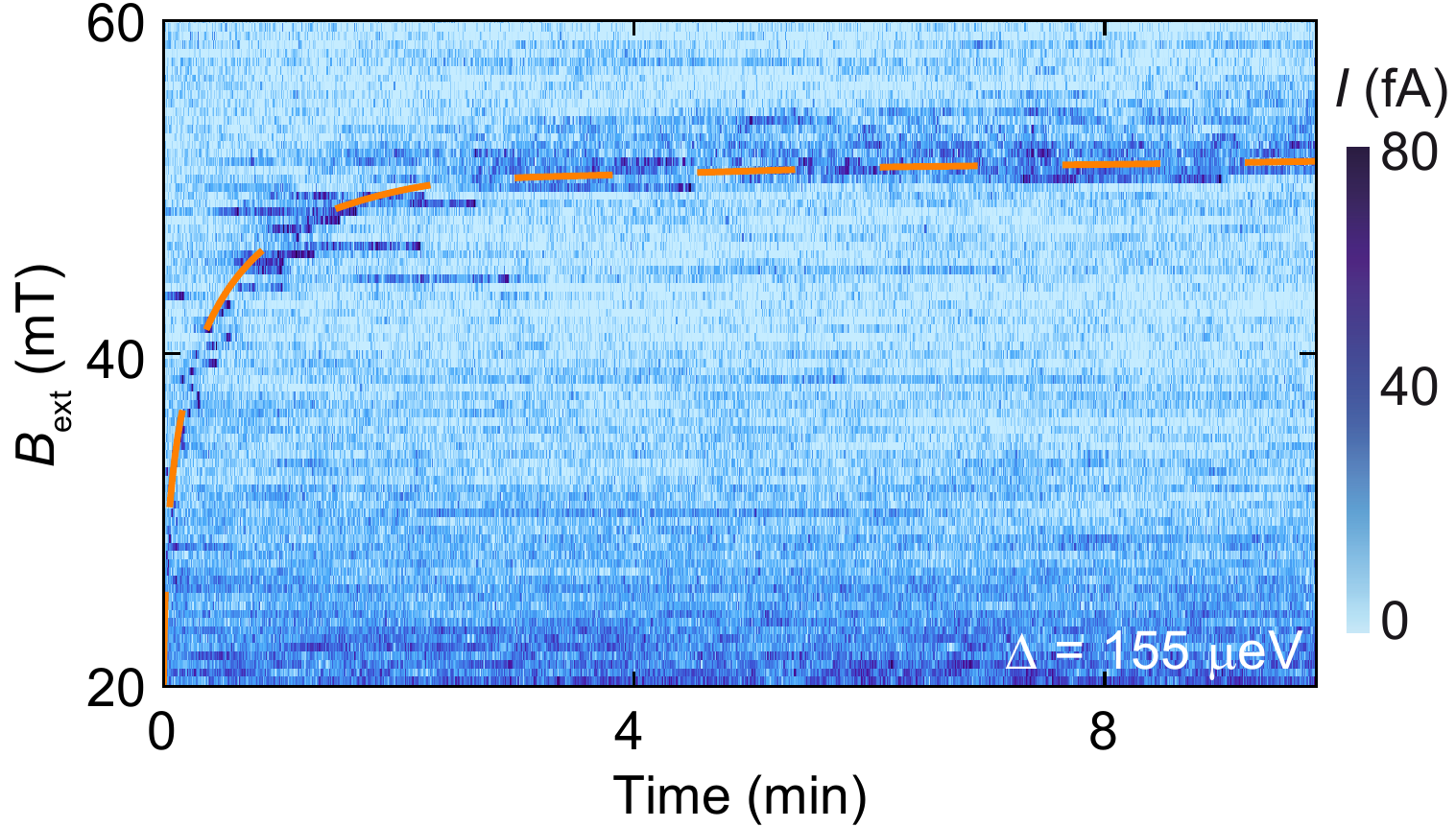}
\caption{
\textbf{Nuclear Spin Dynamics.}
Current measured as a function of time demonstrating DNSP at the $\Sa$--$T_-$ resonance. The dashed orange line is a numerical fit of the data. Here the system is tuned to a larger tunnel coupling compared to the tuning used in Fig.~3 of the main article. As a result, $\tc = \SI{22}{\micro\electronvolt}$ is larger, which makes $E_\text{a}$ larger, and the current peak is not observed until $B_\text{ext} \approx \SI{20}{\milli\tesla}$. Meanwhile, $\alpha = \SI{14}{\per\nano\ampere\per\second}$ suffers from the increased $\tc$. The other fit parameters are unchanged, that is, $\gamma = \SI{10}{\milli\tesla}$ and $\Grel = \SI{0.04}{\per\second}$.
\label{fig:B vs Time}
}
\end{figure}
\clearpage

\end{document}